\documentclass[3p,12pt]{elsarticle}
\pdfoutput=1
\usepackage{amssymb,amsmath,graphicx,hyperref,color,bm,units}

\numberwithin{equation}{section}

\graphicspath{{fig/}}

\DeclareMathOperator{\Tr}{Tr}

\journal{Nuclear Physics B}

\makeatletter
\def\@author#1{\g@addto@macro\elsauthors{\normalsize%
    \def\baselinestretch{1}%
    \upshape\authorsep#1\unskip\textsuperscript{%
      \ifx\@fnmark\@empty\else\unskip\sep\@fnmark\let\sep=,\fi
      \ifx\@corref\@empty\else\unskip\sep\@corref\let\sep=,\fi
      }%
    \def\authorsep{\space {\large and}\space}%
    \global\let\@fnmark\@empty
    \global\let\@corref\@empty
    \global\let\sep\@empty}%
    \@eadauthor={#1}
}

\def\@@author[#1]#2{\g@addto@macro\elsauthors{%
    \def\baselinestretch{1}%
    \authorsep#2\unskip\textsuperscript{
      \@for\@@affmark:=#1\do{%
       \edef\affnum{\@ifundefined{X@\@@affmark}{1}{\elsRef{\@@affmark}}}%
     \unskip\sep\affnum\let\sep=,}%
      \ifx\@fnmark\@empty\else\unskip\sep\@fnmark\let\sep=,\fi
      \ifx\@corref\@empty\else\unskip\sep\@corref\let\sep=,\fi
      }%
    \def\authorsep{\space {\large and}\space}%
    \global\let\sep\@empty\global\let\@corref\@empty
    \global\let\@fnmark\@empty}%
    \@eadauthor={#2}
}

\def\ps@pprintTitle{%
 \let\@oddhead\@empty
 \let\@evenhead\@empty
 \def\@oddfoot{}%
 \let\@evenfoot\@oddfoot}

\long\def\MaketitleBox{%
  \resetTitleCounters
  \def\baselinestretch{1}%
  \begin{center}%
   \def\baselinestretch{1}%
    \Large\@title\par\vskip18pt
    \normalsize\elsauthors\par\vskip10pt
    \footnotesize\itshape\elsaddress\par\vskip36pt
    \rule{\textwidth}{1.5pt}\vskip12pt
    \ifvoid\absbox\else\unvbox\absbox\par\vskip10pt\fi
    \ifvoid\keybox\else\unvbox\keybox\par\vskip10pt\fi
    \rule{\textwidth}{1.5pt}\vskip12pt
    \end{center}%
  }

\renewcommand\subsection{\@startsection{subsection}{2}{\z@}%
           {18\p@ \@plus 6\p@ \@minus 3\p@}%
           {9\p@ \@plus 6\p@ \@minus 3\p@}%
           {\normalfont\normalsize\itshape\bfseries}}

\gdef\emailauthor#1#2{\stepcounter{ead}%
     \g@addto@macro\@elseads{\raggedright%
      \let\corref\@gobble
      \eadsep\newline\texttt{#1} (#2)\def\eadsep{\unskip,\space}}%
}

\renewcommand\appendix{\par
  \setcounter{section}{0}%
  \setcounter{subsection}{0}%
  \setcounter{equation}{0}
  \gdef\thefigure{\arabic{figure}}%
  \gdef\thetable{\@Alph\c@section.\arabic{table}}%
  \gdef\thesection{\appendixname~\@Alph\c@section}%
  \@addtoreset{equation}{section}%
  \gdef\theequation{\@Alph\c@section.\arabic{equation}}%
  \addtocontents{toc}{\string\let\string\numberline\string\tmptocnumberline}{}{}
}

\def\appendixname{Appendix}
\renewcommand\@makefntext[1]{#1}

\makeatother

\biboptions{numbers,sort&compress}

\begin{document}

\begin{frontmatter}

${}$
\vspace{-2cm}
\begin{flushright}
MAN/HEP/2017/02,
ULB-TH/17-03\\[1mm]
March 2017
\end{flushright}
\medskip
\title{{\bf {\LARGE Exact RG Invariance and}}\\[3mm]
{\bf {\LARGE Symmetry Improved 2PI Effective Potential}}\bigskip}

\author[uom]{{\large Apostolos Pilaftsis}}
\author[ulb]{{\large Daniele Teresi}}

\address[uom]{\smallskip Consortium for Fundamental Physics, School of
  Physics and Astronomy,\\ University of Manchester, Manchester M13
  9PL, United Kingdom\\[1mm]
{\rm E-mail address:} {\tt apostolos.pilaftsis@manchester.ac.uk}\bigskip} 
  
\address[ulb]{\smallskip  Service de Physique Th\'eorique, Universit\'e Libre de Bruxelles,\\
Boulevard du Triomphe, CP225, 1050 Brussels, Belgium\\[1mm]
{\rm E-mail address:} {\tt daniele.teresi@ulb.ac.be}}

\begin{abstract}
  The Symmetry Improved Two-Particle-Irreducible (SI2PI) formalism is a powerful tool to calculate the effective potential beyond perturbation theory, whereby infinite sets of selective loop-graph topologies can be resummed in a systematic and consistent manner. In this paper we study the Renormalization-Group (RG) properties of this formalism, by proving for the first time a number of new field-theoretic results. First, the RG runnings of all proper 2PI couplings are found to be UV finite, in the Hartree--Fock and sunset approximations of the 2PI effective action. Second, the SI2PI effective potential is \emph{exactly} RG invariant, in contrast to what happens in the ordinary One-Particle-Irreducible (1PI) perturbation theory, where the effective potential is RG invariant only up to higher orders. Finally, we show how the effective potential of an $\mathbb{O}(2)$ theory evaluated in the SI2PI framework, appropriately RG improved, can reach a higher level of accuracy, even up to one order of magnitude, with respect to the corresponding one obtained in the 1PI~formalism.

\bigskip

\end{abstract}


\end{frontmatter}

\vfill\eject

\makeatletter
\renewcommand\@makefntext[1]{\leftskip=0em\hskip1em\@makefnmark\space #1}
\makeatother

\section{Introduction}
The effective potential constitutes a fundamental tool in Quantum Field Theory, widely used to study a multitude of physical phenomena, such as spontaneous symmetry breaking, tunnelling rates due to vacuum instability, and thermal phase transitions. With the ever increasing experimental precision on the determination of the Standard Model (SM) Higgs boson mass at the CERN Large Hadron Collider (LHC), more accurate computations of the SM effective potential can now be performed up to next-to-next-to-leading order (NNLO). Assuming~no~new physics and ignoring quantum-gravity effects, such NNLO studies~\cite{Bezrukov:2012sa,Degrassi:2012ry,Buttazzo:2013uya} imply that the SM vacuum may well be metastable. Most remarkably, the actual profile of the effective potential at large Higgs-field values was found to be extremely sensitive to very small variations of the input parameters at the electroweak scale~\cite{Degrassi:2012ry,Bezrukov:2014ina}, thereby rendering such precision computations rather subtle. Interestingly enough, this unusual feature seems to have sparked in the last years renewed interest in studying several field-theoretical aspects of the effective potential, which include the development of new semi-analytical techniques beyond ordinary perturbation theory~\cite{Pilaftsis:2013xna,Garbrecht:2015cla,Ellis:2015xwp}, the treatment of infrared (IR) divergences due to the masslessness of the SM would-be Goldstone bosons~\cite{Martin:2014bca,Elias-Miro:2014pca,Pilaftsis:2015bbs}, more precise field-theoretical approaches to estimating tunnelling rates~\cite{Garbrecht:2015oea,Garbrecht:2015yza} and assessing their gauge dependence~\cite{Plascencia:2015pga,Espinosa:2016uaw,Espinosa:2016nld}, as well as evaluations of SM metastability rates in the presence of Planck-scale suppressed operators~\cite{Branchina:2013jra,Branchina:2014rva}.

The Two-Particle-Irreducible (2PI) effective action, as introduced by Cornwall, Jackiw and Tomboulis~\cite{Cornwall:1974vz}, provides a powerful approach to resumming infinite series of higher-order diagrams of a given loop-graph topology in a systematic and self-consistent way.  Nevertheless, simple restrictions of the 2PI effective action to considering only a finite number of loop-graph topologies usually cause residual violations of possible underlying symmetries of the theory, which in turn introduce inconsistencies to the equations of motion. Most notably, for theories realizing spontaneous breakdown of global symmetries, naive truncations of the 2PI effective action lead to massive Goldstone bosons (see, e.g.~\cite{Petropoulos:1998gt}). A potentially interesting theoretical framework for addressing this problem is the so-called Symmetry Improved Two-Particle-Irreducible~(SI2PI) formalism~\cite{Pilaftsis:2013xna}, in which symmetry-consistent restrictions to the 2PI truncated equations of motion were implemented. The resulting SI2PI effective action has a number of satisfactory field-theoretical properties~\cite{Pilaftsis:2013xna}, including the masslessness of the Goldstone bosons and the proper prediction of a second order thermal phase transition for $\mathbb{O}(N)$ scalar theories. Further applications and developments of the SIP2PI formalism may be found in \cite{Pilaftsis:2015cka,Pilaftsis:2015bbs,Mao:2013gva,Brown:2015xma,Garbrecht:2015cla,Brown:2016sak,Marko:2016wtw,Brown:2016vaj}. In particular, in \cite{Pilaftsis:2015bbs} (see also~\cite{Pilaftsis:2015cka}), the effect of fermions has been included in a semi-perturbative way and the SI2PI effective potential has been used to address from {\em first principles} the IR divergence problem of the perturbative SM effective potential due to the masslessness of the would-be Goldstone bosons. Hence, the results of this SI2PI approach have proved to be very useful, as they enable one to assess the regime of validity of other simplified and partially {\it ad hoc} approaches~\cite{Martin:2014bca, Elias-Miro:2014pca}.

In this paper we study the renormalization-group (RG) properties of the SI2PI formalism, and prove for the first time a number of remarkable field-theoretical results. First, after briefly reviewing the formalism in Sec.~\ref{sec:SI2PI}, we show in Sec.~\ref{sec:finite} that the runnings of all proper 2PI quartic couplings are ultra-violet (UV) finite, without ignoring higher-order terms, in the Hartree--Fock and sunset approximations of the 2PI effective action. In fact, as this result is highly non-trivial, one may even conjecture that the UV finiteness of all proper 2PI quartic couplings will hold true to any arbitrarily high loop order of truncation of the SI2PI effective action.

An equally important result of our study is presented in Sec.~\ref{sec:RG_inv}, where we show how the SI2PI effective potential is \emph{exactly} RG invariant at a given loop order of truncation, in stark contrast to what happens in the usual 1PI formalism, where the perturbative effective potential is RG invariant only \emph{up to higher order terms}. This is a remarkable result, as it helps us to clarify the relation between the 2PI resummation and the frequently considered RG-improved approach to the 1PI effective potential.  Specifically, the former takes into account precisely the higher-order contributions needed to restore exact RG invariance when the effective potential is evaluated at a given loop level, whereas the latter attempts to minimize these higher-order terms by plausibly guessing the value of the RG scale~$\mu$, so as to avoid large logarithmic contributions.  At this point, it is important to clarify that the approach we follow here differs conceptually from the one presented in~\cite{Carrington:2014lba,Pawlowski:2015mlf}, where functional methods were employed to reinforce RG invariance in the 2PI effective action.

In Sec.~\ref{sec:RGSI2PI} we modify the SI2PI formalism by appropriately improving the RG running of the quartic couplings. This enables us to reach a high level of accuracy, and compare our results with the 1PI NNLO potential, which includes two-loop threshold corrections matched with 3-loop RG running. We show that, as a consequence of the exact RG-invariance shown in Sec.~\ref{sec:RG_inv}, the predictions for the profile of the effective potential in the SI2PI framework can reach a higher level of accuracy, even up to one order of magnitude, than the corresponding one obtained in the 1PI formalism. This is indeed a significantly new result obtained for a simple $\mathbb{O}(2)$ theory that certainly encourages further studies within the context of more realistic theories.
In a similar vein, in Sec.~\ref{sec:conclusions}  we draw our conclusions and present possible future directions to this field.

\section{Symmetry Improved 2PI Effective Potential}\label{sec:SI2PI}

In this section we briefly review the 2PI formalism by applying it to the scalar $\mathbb{O}(2)$ model and present the equations of motion used to analyze the RG properties derived from this formalism. A more detailed discussion may be found in \cite{Pilaftsis:2013xna, Pilaftsis:2015cka}.

Our starting point is the Lagrangian of the $\mathbb{O}(2)$ model,
\begin{equation}\label{eq:lagr_scal}
\mathcal{L}\ = \ \frac{1}{2} (\partial_\mu \phi^i)(\partial^\mu \phi^i) \;+\; \frac{m^2}{2} (\phi^i)^2 \;-\; \frac{\lambda}{4} (\phi^i)^2 (\phi^j)^2\;,
\end{equation}
where $\phi^i$, $i=1,2$ is the $\mathbb{O}(2)$ scalar multiplet, which can be expanded in terms of a Higgs and a Goldstone mode as
\begin{equation}
\phi^1 \ = \ \phi \,+ \, H \;, \qquad \phi^2 \ = \ G \;.
\end{equation}

The 2PI  effective action is obtained by introducing, in addition to a local source $J(x)$ as in the usual 1PI effective action, a bi-local source $K(x,y)$, both with implicit $\mathbb{O}(2)$ group structure. By Legendre-transforming the connected generating functional with respect to these sources, one obtains the 2PI effective action $\Gamma[\phi,\Delta]$, depending on the background field $\phi$ and the \emph{dressed} propagators $\Delta$. By expanding $\Gamma[\phi,\Delta]$ up to two-loop graph topologies, one finds~\cite{Pilaftsis:2013xna, Pilaftsis:2015cka}
\begin{align}
\label{eq:gamma_scal}
\Gamma^{(2)}[\phi,\Delta_H,\Delta_G] \ =\ & \int \! \bigg[\frac{Z_0}{2} (\partial_\mu \phi)^2 \:+\:\frac{m^2 +
  \delta m_0^2}{2} \,\phi^2 \: -\: 
\frac{\lambda + \delta \lambda_0}{4}  \,\phi^4 \bigg] \notag \\[3pt]
& -\: \frac{i \hbar}{2} \Tr \big(\ln\Delta_H\big) \: -\: 
\frac{i \hbar}{2} \Tr \big(\ln \Delta_G \big) \displaybreak[0]\notag\\[3pt]
&- \: \frac{i \hbar}{2} \Tr \Big\{\Big[ Z_1 \, \partial^2\: +\: 
\big( 3 \lambda + \delta \lambda_1^A + 2 \delta \lambda_1^B\big)\,\phi^2
- \big( m^2 + \delta m^2_1\big) \Big]\,\Delta_H \Big\}  \notag\\[3pt]
&- \: \frac{i \hbar}{2} \Tr \Big\{\Big[ Z_1 \, \partial^2\: +\:
  \big(\lambda + \delta \lambda_1^A\big) \, \phi^2 - \big(m^2 + \delta m^2_1\big)
  \Big]\,\Delta_G \Big\}  \displaybreak[0]  \notag\\[3pt]
  & - \, \frac{i \hbar^2}{4} \, \bigg\{ -i(3 \lambda + \delta \lambda_2^A + 2 \delta \lambda_2^B) \int \! i\Delta_H i\Delta_H  \;-\; 2i (\lambda + \delta \lambda_2^A ) \int \! i\Delta_H i\Delta_G\notag\\ 
  &-\; i (3\lambda + \delta \lambda_2^A + 2 \delta \lambda_2^B) \int \! i\Delta_G i\Delta_G \bigg\} \notag\\ 
  & - \, i \hbar^2 \, \Bigg\{ \; \parbox{1.7cm}{\includegraphics[width=1.7cm]{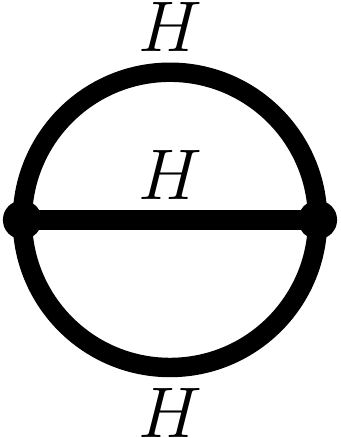}} \; + \; \parbox{1.7cm}{\includegraphics[width=1.7cm]{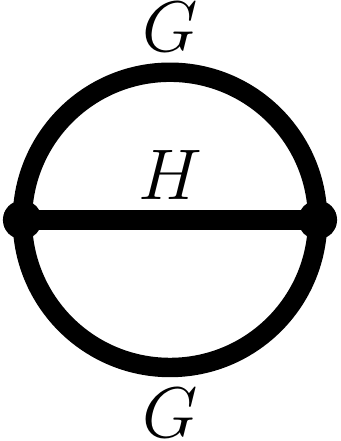}} \; \Bigg\} \;,
\end{align}
where the integrals are meant to be evaluated in position space over the common spacetime variable of the relevant fields and propagators. In the last line of~\eqref{eq:gamma_scal}, thick lines denote the dressed Green's functions $\Delta$, whereas in the following thin lines are reserved to represent tree-level propagators.

The additional counter-terms (CTs) in~\eqref{eq:gamma_scal} with respect to the 1PI formalism are related to the appearance of several operators with mass-dimensions 2 and 4 in the 2PI effective action~\cite{vanHees:2001ik,Blaizot:2003an,Berges:2005hc}. In addition, there is a perturbative 1PI-like CT $\delta \lambda$ for the quartic coupling $\lambda$ that appears at two- and higher loop orders~\cite{Berges:2005hc}. At the order of truncation considered here, however, this last CT~$\delta\lambda$ is not needed to be included in the vertices of the diagrams in the last line of~\eqref{eq:gamma_scal}, as such a CT would be necessary to cancel subdivergences of higher-order diagrams, in close analogy to the usual 1PI perturbation theory. 

For the \emph{full} effective action $\Gamma[\phi,\Delta]$, the Equation of Motions (EoMs) are given by
\begin{equation}\label{eq:diff}
\frac{\delta \Gamma}{\delta \phi} \, = \, 0 \;, \qquad \frac{\delta \Gamma}{\delta \Delta} \, = \, 0\;. 
\end{equation}
In practice, however, the effective action can be calculated by considering only a finite number of 2PI topologies, for instance up to two-loop graph topologies as in~\eqref{eq:gamma_scal}, so that we can write $\Gamma = \Gamma^{(2)} + \Gamma_{?}$, where $\Gamma_{?}$ is unknown. A common strategy is to set $\Gamma_{?}=0$, based on our ignorance of it, so that the EoMs are obtained by replacing $\Gamma \to \Gamma^{(2)}$ in~\eqref{eq:diff}. However, in the presence of symmetries this leads to a number of artefacts, most importantly to violations of the usual Ward identities, thus giving rise to massive Goldstone bosons (see~\cite{Pilaftsis:2013xna} and references therein) in
theories with spontaneous breaking of global symmetries. This long-standing issue was addressed in~\cite{Pilaftsis:2013xna} (see also \cite{Pilaftsis:2015bbs}) by considering the so-called Symmetry Improved Two-Particle Irreducible (SI2PI) formalism. In this approach, rather than setting $\Gamma_{?}=0$, one exploits the symmetries of the system to improve upon the expected form 
of $\Gamma_{?}$, implying that the EoMs needed to be solved are
\begin{equation}\label{eq:SI2PI}
\frac{\delta \Gamma^{(2)}}{\delta \Delta} \, = \, 0 \;, \qquad  -\phi \, \Delta_G^{-1}(k=0) \, = \, 0 \;.
\end{equation}
Notice that, although we have written the above equations using the two-loop order truncation of the 2PI effective action in~\eqref{eq:gamma_scal}, these will still be valid at any order of truncation. 
It has been shown in~\cite{Pilaftsis:2013xna}  that this approach has a number of satisfactory field-theoretical properties, besides the masslessness of the Goldstone bosons in the spontaneously-symmetry-broken phase of the theory. 
Moreover, the SI2PI effective potential $V(\phi)$ may be uniquely determined as a solution to the first-order differential equation
\begin{equation}
   \label{eq:SI2PIpotential}
\frac{d V}{d \phi} \ = \ -  \phi \, \Delta_G^{-1}(k=0, \phi) \;,
\end{equation}
subject to the initial condition $V(\phi = v ) = 0$, where $v$ is the VEV of the scalar field $\phi^1$.

As discussed above, the EoMs for the dressed propagators~$\Delta_{H,\,G}(k)$ are obtained by differentiating~$\Gamma^{(2)}[\phi,\Delta_H,\Delta_G]$ in~\eqref{eq:gamma_scal} with respect to these propagators. After Wick-rotating to the Euclidean momentum space, we obtain 
\begin{align}
  \Delta^{-1}_H(k)\ =&\  \ (1 + \delta Z_1) \, k^2 \:+\: (3 \lambda + \delta \lambda_1^A +
                         2\delta\lambda_1^B)\, \phi^2 \:-\: (m^2 + \delta m_1^2)  \notag\\[3pt]  &\ +\: (3
                                                                                                   \lambda + \delta\lambda_2^A + 2 \delta \lambda_2^B) \, \hbar \bm{\mathcal{T}}_{\!\!H} \: +\:  (\lambda + \delta \lambda_2^A) \, \hbar \bm{\mathcal{T}}_{\!\!G} \notag\\[3pt] &\ - \:
                                                                                                   18 \lambda^2 \phi^2 \, \hbar \bm{\mathcal{I}}_{HH}(k) \:-\: 2 \lambda^2 \phi^2 \, \hbar \bm{\mathcal{I}}_{GG}(k) \;, 
  \label{eq:eombareHscal}                                                                                                  \displaybreak[0]\\[9pt]
  \Delta^{-1}_G(k)\  =&\ \ (1 + \delta Z_1) \, k^2 \:+\: (\lambda + \delta \lambda_1^A )\, \phi^2 \:-\: (m^2 + \delta m_1^2) \notag\\[3pt]  &\ +\: (\lambda + \delta \lambda_2^A) \, \hbar\bm{\mathcal{T}}_{\!\!H} 
 \:+\:  (3
\lambda +  \delta\lambda_2^A + 2 \delta \lambda_2^B)\,\hbar \bm{\mathcal{T}}_{\!\!G} \: - \:
4 \lambda^2 \phi^2 \, \hbar \bm{\mathcal{I}}_{HG}(k) \;, \label{eq:eombareGscal}
\end{align}
where $\delta Z_1 = Z_1 -1$. Here, we have introduced the tadpole and sunset integrals:
\begin{equation}\label{eq:TaIab}
\bm{\mathcal{T}}_{\!\!a} \ =\ \overline{\mu}^{2\epsilon} \int_p i \Delta_a(p)
\;, \qquad\qquad  
\bm{\mathcal{I}}_{ab}(k) \ =\ \overline{\mu}^{2\epsilon} \int_p
i \Delta_a(k + p) \, i \Delta_b(p) \;, 
\end{equation}
where $a,b = H,G$ and $\ln\overline{\mu}^2 = \ln \mu^2 + \gamma - \ln(4\pi)$, and $\mu$ is the so-called $\overline{\rm MS}$ renormalization mass scale. 

At the two-loop level of the 2PI effective action, there is no wavefunction re\-normalization for the radial field~$\phi$, and so we can set its CT $\delta Z_1$ to zero. The~EoMs~\eqref{eq:eombareHscal} and \eqref{eq:eombareGscal} are renormalized by cancelling separately the subdivergences proportional to the re\-normalized tadpole integrals and the overall divergences proportional to the field powers~$\phi^0$ and~$\phi^2$, as described in detail in \cite{Pilaftsis:2013xna,Fejos:2007ec,Pilaftsis:2015bbs}. Out of $2 \times 4$ relations, only 5 of them are found to be independent, which uniquely fixes the value of the 5 CTs appearing in~\eqref{eq:eombareHscal} and \eqref{eq:eombareGscal}. Following this procedure, the renormalized EoMs are found to be~\cite{Pilaftsis:2013xna,Pilaftsis:2015cka} 
 \begin{align}
  \Delta^{-1}_H(k) \ =\ &\ k^2 \:+\: (\lambda_1^A + 2 \lambda_1^B) \phi^2 \,-\, m^2  \:+\: (\lambda_2^A + 2 \lambda_2^B) \, \hbar \mathcal{T}^{\rm fin}_H \:+\: \lambda_2^A \,\hbar \mathcal{T}_G^{\rm fin} \: - \:
 18\lambda^2\phi^2 \, \hbar \mathcal{I}_{HH}^{\rm fin}(k) \notag\\
 &- \: 2\lambda^2\phi^2 \, \hbar \mathcal{I}_{GG}^{\rm fin}(k) \;+\; \hbar^2 \Pi^{\mathrm{2PI}, (2)}_H \;,   \label{eq:eomren_H} \\[6pt]
\Delta^{-1}_G(k) \ =\ & \ k^2 \:+\: \lambda_1^A \phi^2 \,-\, m^2  \:+\: \lambda_2^A  \, \hbar\mathcal{T}_H^{\rm fin} \:+\: (\lambda_2^A + 2 \lambda_2^B)  \,\hbar \mathcal{T}_G^{\rm fin} \: - \:
 4\lambda^2\phi^2 \, \hbar \mathcal{I}_{HG}^{\rm fin}(k) \notag\\
&+ \: \hbar^2\Pi^{\mathrm{2PI}, (2)}_G  \;, \label{eq:eomren_G}
\end{align}
where the analytic expressions for the renormalized, UV-finite integrals $\mathcal{T}_a^{\rm fin}$ and $\mathcal{I}_{ab}^{\rm fin}$ are given in \ref{app:sunset}. Keeping only the tadpole integrals $\mathcal{T}_a^{\rm fin}$ in \eqref{eq:eomren_H} and \eqref{eq:eomren_G} is known as the Hartree--Fock (HF) approximation. Including also the remaining one-loop integrals $\mathcal{I}_{ab}^{\rm fin}$ is commonly referred to as the sunset approximation. In order to obtain a level of accuracy higher than the two-loop order in perturbation theory, we have included also the renormalized two-loop 2PI self-energies $\Pi^{\mathrm{2PI}, (2)}_a$. However, as discussed in~\cite{Pilaftsis:2015cka,
Pilaftsis:2015bbs}, the latter contributions result from a three-loop order truncation of the 2PI effective action, and so we approximate them by their usual 1PI form evaluated in the zero-momentum limit $k\to 0$. Their analytic expressions are given in~\ref{app:pert_2loop}. We will include them only in Sec.~\ref{sec:RGSI2PI} when calculating explicitly the SI2PI effective potential. In~\eqref{eq:eomren_H} and \eqref{eq:eomren_G} we have generalized the analysis in~\cite{Pilaftsis:2013xna} by taking the various 2PI quartic couplings different from each other, even at the renormalized level (not only the corresponding CTs). This is done so in anticipation of the results that will be derived in the next section, 
where we will show that the RG running of the various 2PI quartic couplings will be, in general, different and so they can be taken to be equal only at some fixed RG scale $\mu^*$.

\begin{figure}
\centering
\includegraphics[width=0.9\textwidth]{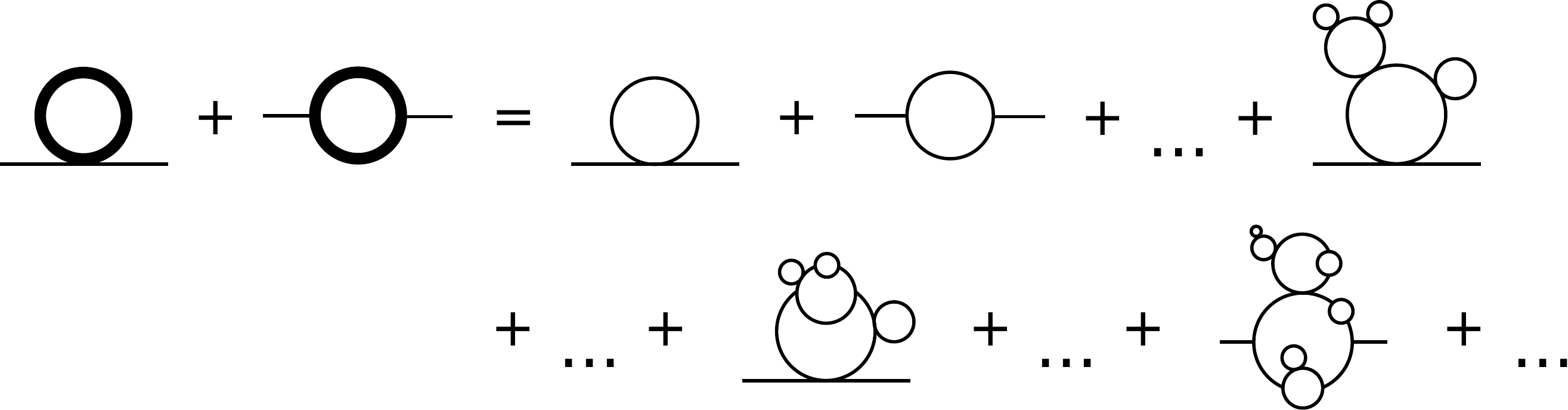}
\caption{ Typical topologies of graphs that are implicitly resummed by the one-loop 2PI self-energies, i.e.~\eqref{eq:eomren_H} and~\eqref{eq:eomren_G} in the sunset approximation. 
\label{fig:resum1loop}}
\end{figure}

\begin{figure}
\centering
\includegraphics[width=0.8\textwidth]{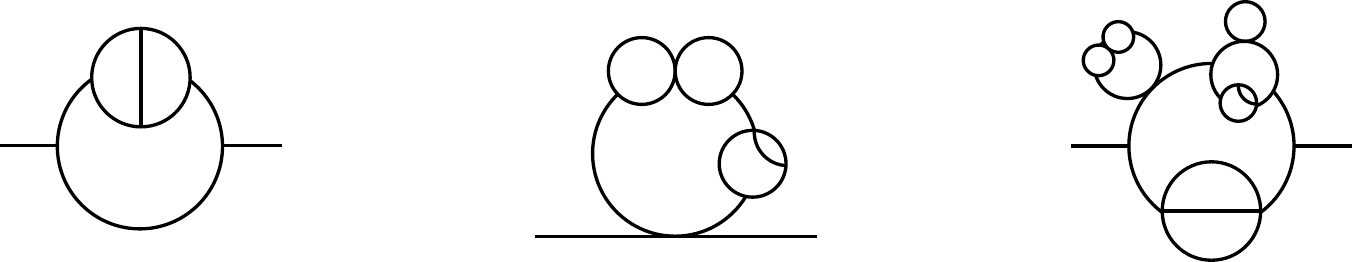}
\caption{Typical set of graphs that are also resummed when including the two-loop 2PI self-energies $\Pi^{\mathrm{2PI}, (2)}_a $ in the EoMs~\eqref{eq:eomren_H} and~\eqref{eq:eomren_G}. Notice that the propagators belonging to two-loop 2PI topologies do not
get dressed, since these are approximated by their usual 1PI forms.\label{fig:resum2loop}}
\end{figure}

As shown in Figs.~\ref{fig:resum1loop} and \ref{fig:resum2loop}, the EoMs \eqref{eq:eomren_H} and \eqref{eq:eomren_G} represent selective resummations of infinite set of diagrams of certain loop-graph topologies, which are obtained by dressing the propagators with one- and two-loop self-energies by an arbitrary number of times. Here, we must note that the one-loop self-energies can also be dressed, in a recursive way by an arbitrary number of times in fair analogy to considerations from fractal geometry, whereas the two-loop self-energies cannot, since these have been approximated by their usual 1PI forms.

Before concluding this section, we briefly comment  on the potential issue regarding the IR sensitivity of the sunset approximation in the SI2PI formalism, which was recently raised in \cite{Marko:2016wtw}, concerning the (IR-divergent) Higgs propagator in the limit $k \to 0$. We have checked explicitly that for values of scalar quartic couplings under study, i.e.~$\lambda \sim 0.13$, the alleged IR-sensitive effects on IR-safe observables of interest here and in~\cite{Pilaftsis:2013xna}, such as the pole mass of the unstable Higgs particle and the 2PI effective potential, are absent at the level of accuracy we have been considering. In fact, we have numerically varied the IR cutoff by a few orders of magnitude and carefully verified that our results for the effective potential remain stable. This  check for the effective potential away from its minimum complements analogous checks that were performed already in \cite{Pilaftsis:2013xna} at the minimum of the potential.

\section{Finiteness of the RG Running}\label{sec:finite}

In this section we calculate the RG running of the proper 2PI couplings in the HF and sunset approximations, and prove that they are UV finite without ignoring higher order terms, as usually done in the 1PI perturbation theory. This is the first important result of this analysis, which holds true for both the 2PI and  SI2PI formalisms. In fact, the UV finiteness of the proper 2PI couplings is a highly non-trivial RG property.  As we illustrate in Sec.~\ref{sec:1PI_finite}, it requires delicate cancellations between $1/\epsilon^n$ poles with different $n$-powers and at different orders in the loop ($\hbar$) expansion. While in 1PI perturbation theory this result is guaranteed by ``$\hbar$-expanding'' the full 1PI effective action, no analogous argument holds for the respective 2PI action, since no series-expansion parameter exists in this case for the solutions to the EoMs.

\subsection{Finiteness of the 1PI RG Running}\label{sec:1PI_finite}

Following~\cite{Machacek:1983tz}, the bare quartic couplings $\lambda_i^{\rm bare}$ assume in the 1PI $\overline{\rm MS}$ scheme the general form 
\begin{equation}\label{eq:1PI_exp}
\lambda_i^{\rm bare} \ = \ \overline{\mu}^{2 \epsilon} \, ( \lambda_i  + \delta \lambda_i ) \ = \ \overline{\mu}^{2 \epsilon} \, \bigg( \lambda_i \, +\, \sum_{n=1}^{\infty} \frac{a^i_{n}}{\epsilon^n} \bigg)
\end{equation}
in $d= 4 - 2 \epsilon$ dimensions, where $\epsilon$ is a dimensional-regularization (DR) parameter. Upon imposition of $\mu$-independence on the bare quartic couplings in~\eqref{eq:1PI_exp}, 
\begin{equation} \label{eq:bare_dont_run}
\mu \, \frac{d}{d \mu} \, \lambda_i^{\rm bare} \ = \ 0 \;,
\end{equation}
one finds that the RG running of the renormalized quartic couplings $\lambda_i$ depends only on the coefficients of the single poles $a_1^i$, i.e.
\begin{equation}
   \label{eq:beta_running}
\mu \, \frac{d \lambda_i}{d \mu} \ = \ - 2 \epsilon \lambda_i \, + \, \beta_i \; , 
\end{equation}
with
\begin{equation}
   \label{eq:beta_i}
\beta_i \ = \ - 2 a_1^i  +  \sum_j 2 \lambda_j \frac{\partial a_1^i}{\partial \lambda_j} \;.
\end{equation}
Instead, the coefficients $a_n$ of the higher-order $1/\epsilon^n$ poles, with $n \geq 2$, have to conspire, in such a way that the RHS of \eqref{eq:bare_dont_run} vanishes identically. This can happen, {\it iff} the recursive relations, 
\begin{equation}\label{eq:recursion}
2 a_{n+1}^i \ = \ \sum_j 2 \lambda_j \frac{\partial a^i_{n+1}}{\partial \lambda_j} \, - \, \sum_j \beta_j \frac{\partial a^i_n}{\partial \lambda_j}\; ,
\end{equation}
are satisfied for all $n \geq 1$. We note that these recursive relations are a consequence of subtle cancellations
required in order to have finite $\beta$-functions $\beta_i$. Most remarkably, they involve coefficients $a_n$ of poles with different powers of $\epsilon$, and when \eqref{eq:beta_running} and \eqref{eq:recursion} are expanded in series of loops, these cancellations occur among different loop levels.  

All the relations presented in this subsection are valid to all orders in perturbation theory. In the 1PI formalism, they can be expanded in a series of the loop parameter $\hbar$, and are therefore valid at a given loop order as well. This result shows the UV finiteness of the RG running at fixed loop order in the 1PI formalism.

\subsection{2PI RG Running}

We first consider the RG running of all 2PI couplings in the HF approximation, which will 
help us to gain valuable insight into the structure of the CTs and their RG properties. Then, we 
extend our considerations to the less trivial case of the sunset approximation.

In order to renormalize the EoMs in the HF approximation (cf.~\eqref{eq:eombareHscal} and \eqref{eq:eombareGscal} with the sunset integrals $\bm{\mathcal I}_{ab}$ neglected) as sketched in Sec.~\ref{sec:SI2PI} and detailed in~\cite{Pilaftsis:2013xna}, the following CTs are needed:
\begin{align}
  \label{eq:dl1PI}
\delta \lambda \ &=\ 0\;, \\[2mm]
\delta \lambda_1^A \ =\ \delta \lambda_2^A \ &= \ \frac{L \lambda^A}{\epsilon} \;  \frac{\displaystyle  \lambda^A \bigg(2 - \frac{4 L \lambda^B}{\epsilon} \bigg) \,+\, 4 \lambda^B \bigg( 1 - \frac{L \lambda^B}{\epsilon}\bigg)}{\displaystyle 1 - \frac{2 L (\lambda^A + 2 \lambda^B)}{\epsilon} + \frac{4 L^2 \lambda^B(\lambda^A+\lambda^B)}{\epsilon^2}} \;,\label{eq:dl_A} \\
\delta \lambda_1^B\ =\ \delta \lambda_2^B \ &= \ \frac{L \lambda^B}{\epsilon} \; \frac{2 \lambda^B}{\displaystyle 1- \frac{2 L \lambda^B}{\epsilon}} \;,\label{eq:dl_B}\\
\delta m^2 \ &= \ \frac{L m^2}{\epsilon} \; \frac{2(\lambda^A +\lambda^B)}{\displaystyle 1 - \frac{2 L (\lambda^A + \lambda^B)}{\epsilon}}\;,\label{eq:dm}
\end{align}
with $L \equiv \hbar/(16 \pi^2)$ and $\delta m^2 \equiv \delta m^2_1$. Here, we have generalized the results reported in~\cite{Pilaftsis:2013xna}, by taking $\lambda^A \neq \lambda^B$,
with $\lambda^A = \lambda^A_1 = \lambda^A_2$ and $\lambda^B = \lambda^B_1 = \lambda^B_2$.
As we will see below, we have done so,  since $\lambda^A$ and $\lambda^B$ run differently. 

Notice that the CTs stated in~\eqref{eq:dl1PI}--\eqref{eq:dm} contain contributions from all orders in $L$ or~$\hbar$, including infinite inverse powers of $\epsilon$. Hence, it is far from obvious that these CTs will lead to a finite running for the renormalized couplings $\lambda^A$ and $\lambda^B$, according to our discussion in Sec.~\ref{sec:1PI_finite}. However, as we will now show, the RG 
running of the proper 2PI couplings $\lambda^A$ and $\lambda^B$ is {\em exactly} UV finite, within the HF approximation {\em itself}.

To this end, let us first consider the expansion of~\eqref{eq:dl_B} in powers of $1/\epsilon$,
\begin{equation}
\delta \lambda_{1,2}^B \ = \ \frac{2 L (\lambda^B)^2 }{\epsilon} \, + \, \frac{4 L^2 (\lambda^B)^3 }{\epsilon^2} \, + \, \ldots \;.
\end{equation}
Employing the notation of Sec.~\ref{sec:1PI_finite}, we may identify $a_1 = 2 L (\lambda^B)^2$ and $a_2 = 4 L^2 (\lambda^B)^3$. Hence, \eqref{eq:beta_running} yields $\beta^B = 4 L (\lambda^B)^2$ and the $1/\epsilon$ pole in the running is cancelled, provided \eqref{eq:recursion}, for $n=1$, is satisfied, i.e.
\begin{equation}
8 L^2 (\lambda^B)^3 \ =\ 2 \lambda^B \times 12 L^2 (\lambda^B)^2 \, - \, 4 L (\lambda^B)^2 \times 4 L \lambda^B \;.
\end{equation}
It is not difficult to verify that this is indeed the case. To prove the cancellation of all $1/\epsilon^n$, it is more convenient to follow a different strategy and use \eqref{eq:bare_dont_run} directly with the all-orders expressions~\eqref{eq:dl1PI}--\eqref{eq:dm}. For instance, by plugging~\eqref{eq:dl_B} into~\eqref{eq:bare_dont_run}, we must have
\begin{align}
\mu \frac{d}{d\mu} \Big[\,\overline{\mu}^{2 \epsilon} \Big( \lambda^B + \delta \lambda^B\Big)\Big]\  &= \ \mu \frac{d}{d\mu} \bigg[ \, \overline{\mu}^{2 \epsilon} \bigg( \lambda^B + \frac{2 L (\lambda^B)^2}{\epsilon} \, \frac{1}{ 1- 2 L \lambda^B/\epsilon} \bigg) \bigg] \notag \\
& = \ \overline{\mu}^{2 \epsilon} \frac{2 \epsilon \lambda^B}{ 1- 2 L \lambda^B/\epsilon} \;+\; \overline{\mu}^{2 \epsilon} \frac{1}{(1- 2 L \lambda^B/\epsilon)^2}\, \mu \frac{d \lambda^B}{d \mu}\ =\ 0 \;.
\end{align}
This last constraint implies the finite RG running for $\lambda^B$,
\begin{equation}\label{eq:running_HF_B}
\mu \frac{d }{d \mu}\lambda^B \ = \ - 2 \epsilon \lambda^B \, + \, 4 L (\lambda^B)^2 \;,
\end{equation}
within the HF approximation only. By analogy, the RG running of the remaining parameters in the HF approximation are found to be
\begin{align}
\mu \frac{d }{d \mu} \lambda \ &= \ 0\;, \label{eq:running_HF_lambda}\\
\mu \frac{d }{d \mu} \lambda^A\ &= \ - 2 \epsilon \lambda^A \, + \, 4 L \lambda^A(\lambda^A + 2 \lambda^B) \;, \label{eq:running_HF_A}\\
\mu \frac{d}{d \mu}  m^2 \ &= \ 4 L \, m^2 \,(\lambda^A +\lambda^B) \;.\label{eq:running_HF_m}
\end{align}
From the finiteness of the above RG equations, we see that the RG running is fully determined by 
the single $1/\epsilon$ poles of the CTs, whereas all higher order $1/\epsilon^{n > 1}$ poles occurring in the resummed 2PI CTs in~\eqref{eq:dl1PI}--\eqref{eq:dm} cancel precisely against each other.
Evidently, we observe the extraordinary RG property that {\em higher-order poles in the resummed
2PI CTs appear in such a way that make no contributions to the RG running, in complete analogy with what happens in the 1PI formalism to all orders.} 

\begin{figure}
\centering
\includegraphics[height=0.4\textwidth]{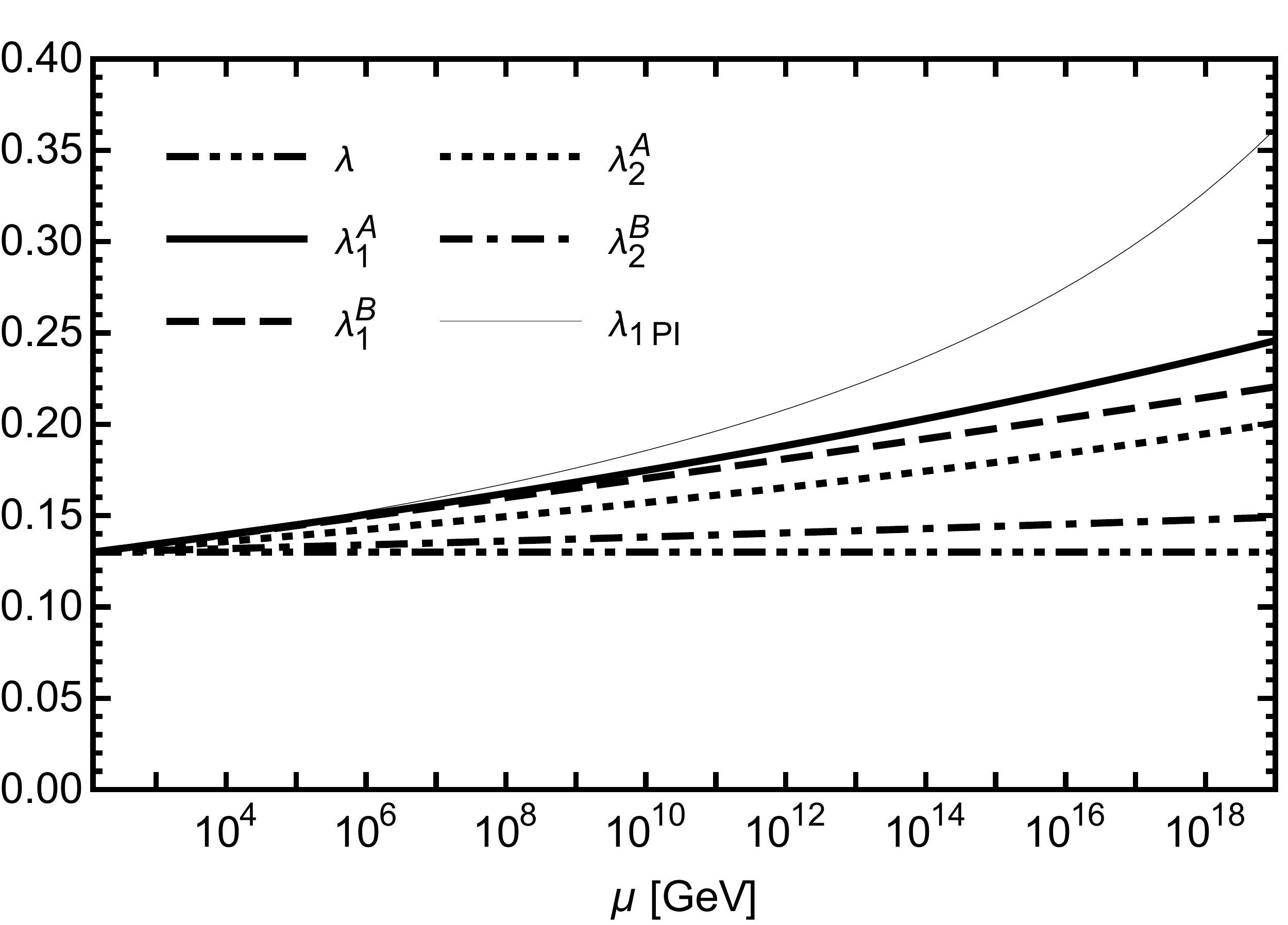}
\caption{The running of the various 2PI quartic couplings in the sunset approximation, starting from a common value $\lambda^* = 0.13$ at $\mu^* = \unit[125]{GeV}$. For comparison, we also show the one-loop running of the coupling $\lambda_{\rm 1PI}$ in the 1PI formalism. \label{fig:running}}
\end{figure}

We now turn our attention to the RG running of the 2PI kinematic parameters in the sunset approximation. Proceeding as above and using the formulae for the CTs reported in~\ref{app:sunset}, 
the following RG equations may be derived:
\begin{align}
\displaystyle \mu \frac{d}{d \mu} \lambda \ &= \ - 2 \epsilon \, \lambda \;,\label{eq:running_sunset_lambda}\\
\displaystyle \mu \frac{d}{d \mu} \lambda_1^A \ &= \ - 2 \epsilon \, \lambda_1^A \,+\, 4 L \, (\lambda_1^B \lambda_2^A + \lambda_1^A \lambda_2^A + \lambda_1^A \lambda_2^B + 2\lambda^2) \,-\, 16 L^2  \lambda^2 (3 \lambda_2^A + \lambda_2^B) \;, \label{eq:running_sunset_l1A}\displaybreak[1]\\
\displaystyle \mu \frac{d}{d \mu} \lambda_1^B \ &= \ - 2 \epsilon \, \lambda_1^B \,+\, 4 L \, (\lambda_1^B \lambda_2^B + 4\lambda^2) \,-\, 32 L^2 \lambda^2 \lambda_2^B \;, 
\label{eq:running_sunset_l1B}\displaybreak[1]\\
\displaystyle \mu \frac{d}{d \mu} \lambda_2^A \ &= \ - 2 \epsilon \, \lambda_2^A \,+\, 4 L \, \lambda_2^A (\lambda_2^A + 2\lambda_2^B) \;, 
\label{eq:running_sunset_l2A}\displaybreak[1]\\
\displaystyle \mu \frac{d}{d \mu} \lambda_2^B \ &= \ - 2 \epsilon \, \lambda_2^B \,+\, 4 L \, (\lambda_2^B)^2\;,\label{eq:running_sunset_l2B} \displaybreak[1]\\
\displaystyle \mu \frac{d}{d \mu} m^2 \ &= \ \,  4 L \, m^2 \, (\lambda_2^A + \lambda_2^B)\;. \label{eq:running_sunset_m}
\end{align}
Notice that the RG equations~\eqref{eq:running_sunset_l1A} and \eqref{eq:running_sunset_l1B} for the proper 2PI couplings $\lambda_1^A$ and $\lambda^B_1$, respectively, contain contributions up to two-loop order. Even though the sunset EoMs are formally expressed in terms of 2PI one-loop integrals only, the resulting $O(L^2)$ terms in the RG runnings of $\lambda_1^A$ and $\lambda^B_1$ are a consequence of non-trivial resummation effects. Consequently, in the sunset approximation all the quartic couplings exhibit different RG runnings.  This is shown in Fig.~\ref{fig:running}, where the RG running of all quartic couplings is displayed, using the common value $\lambda^* = 0.13$ at $\mu^* = \unit[125]{GeV}$ as initial renormalization condition. This choice of values for $\lambda^*$ and  $\mu^*$ was motivated by the expected strength of the SM Higgs quartic coupling at scales close to the mass of the observed Higgs boson at the LHC.  Although the RG runnings of all different quartic couplings will coincide when the full 2PI effective action is considered, we observe that the spread between them is still significant in the sunset approximation.

\section{Exact RG Invariance}\label{sec:RG_inv}

In this section our aim is to prove that  the SI2PI effective potential is {\em exactly} RG invariant in the HF approximation. In ~\ref{app:RG_invariance}, we extend our proof of exact RG invariance to the full one-loop truncated SI2PI effective potential which includes the sunset approximation.

This explicit demonstration of exact RG invariance of the SI2PI effective potential is one of the central results of this work. We should remind ourselves that no exact RG invariance can be achieved in the 
perturbative 1PI effective potential. In fact, the perturbative 1PI effective potential is approximately RG invariant, as the residual RG-violating terms are formally of higher order in the $\hbar$ expansion.
To minimize the significance of these higher order terms, one makes educated guesses for the 
value of the RG scale $\mu$, usually taken to be close to the field value $\phi$, in order to avoid potentially sizeable uncertainties from large logarithmic contributions to the effective potential.
This approach forms the basis of the well-known procedure of RG improvement of the 1PI effective potential. Instead, this approach of RG improvement is no longer necessary within the SI2PI framework. 
As we explicitly show below, the SI2PI effective potential is {\em exactly} RG invariant
by itself, and no {\it ad hoc} choices of the value of the RG scale $\mu$ need  to be made. In particular,
for a fixed value of $\phi$, the SI2PI effective potential does not depend on~$\mu$, so any choice of $\mu$ can be made, including $\mu = \phi$, for example.

Proving the exactness of RG invariance in the HF approximation becomes a straight\-forward exercise, because the integrals involved can be performed analytically. In particular, we have $\Delta_a^{-1}(k) = k^2 + M_a^2$ in the Euclidean space and~\cite{Pilaftsis:2013xna}
\begin{equation}\label{eq:tadpoleHF}
\hbar \mathcal{T}_a^{\rm fin} \ = \ L M_a^2 \bigg( \ln \frac{M_a^2}{\mu^2} - 1 \bigg) \;,
\end{equation}
with $a = H,G$.
Let us now analyze the responses of all relevant parameters under an infinitesimal shift of  the renormalization scale: $t \to t+ \delta t$, with $t\equiv \ln (\mu/\mu^*)$. Obviously, the field $\phi$ remains constant under this shift, i.e.~$\delta \phi = 0$, since the wavefunction renormalization vanishes at this order~\cite{Pilaftsis:2013xna}. The other parameters of the 2PI effective action then vary as follows:
\begin{equation}\label{eq:shift_param}
\delta m^2\ =\  m^2 \gamma_{m^2}\,\delta t\;, \qquad \delta \lambda\ =\ 0\;, \qquad 
\delta \lambda^{A,B}\ =\ \beta_{A,B} \,\delta t \;,
\end{equation}
where $\delta t = \delta \mu/\mu$ and the explicit form of the $\beta$- and $\gamma$-functions can be read off from~\eqref{eq:running_HF_B}--\eqref{eq:running_HF_m} for $\epsilon = 0$. Likewise, the variation of the tadpole integral in~\eqref{eq:tadpoleHF} is given by
\begin{equation}
   \label{eq:delta_tadpoleHF}
\delta (\hbar \mathcal{T}_a^{\rm fin}) \ = \ L  \ln \bigg(\frac{M^2_a}{\mu^2}\bigg)\,\delta M_a^2 \: - \: 2 L M_a^2\,\delta t \;.
\end{equation}
Under the same infinitesimal RG shift: $\mu \to \mu + \delta \mu$, the variation of the EoM~\eqref{eq:eomren_H} for the field $H$ in the HF approximation may be calculated 
by virtue of~\eqref{eq:shift_param} and~\eqref{eq:delta_tadpoleHF} to be
\begin{align}
\delta M_H^2 \ = & \ (\beta_A + 2 \beta_B) \phi^2 \delta t \, - \, m^2 \gamma_{m^2} \delta t \notag\\ 
&+\, (\beta_A + 2 \beta_B) \hbar \mathcal{T}_H^{\rm fin} \delta t \,+ \, (\lambda^A + 2 \lambda^B) \bigg[ L  \ln \bigg(\frac{M^2_H}{\mu^2}\bigg)\delta M_H^2 \, - \, 2 L M_H^2 \delta t \bigg] \notag \\
&+\, \beta_A  \hbar \mathcal{T}_G^{\rm fin} \delta t \,+ \, \lambda^A\bigg[ L  \ln \bigg(\frac{M^2_G}{\mu^2}\bigg)\delta M_G^2 \, - \, 2 L M_G^2 \delta t \bigg] \;.
\end{align}
Making now use of the explicit forms for the $\beta$- and $\gamma$-functions, as given by \eqref{eq:shift_param} and \eqref{eq:running_HF_B}--\eqref{eq:running_HF_m}, we finally arrive at
\begin{equation}
   \label{eq:delta_EoM_H}
\bigg[1 - (\lambda^A+2 \lambda^B) L  \ln \bigg(\frac{M^2_H}{\mu^2}\bigg) \bigg] \delta M_H^2 \, -\, \bigg[\lambda^A L  \ln \bigg(\frac{M^2_G}{\mu^2}\bigg) \bigg] \delta M_G^2  \ = \ 0\times \delta t\ =\ 0\;.
\end{equation}
It is important to remark here that all terms proportional to $\delta t = \delta \mu /\mu$ vanish identically.

We may now proceed in a similar fashion by considering the variation of the EoM~\eqref{eq:eomren_G}  for the Goldstone field $G$ in the HF approximation. In this case, we obtain another homogeneous equation linearly 
independent of~\eqref{eq:delta_EoM_H} for $\delta M_H$ and $\delta M_G$,
\begin{equation}
   \label{eq:delta_EoM_G}
\bigg[\lambda^A L  \ln \bigg(\frac{M^2_H}{\mu^2}\bigg) \bigg] \delta M_H^2\, -\,
\bigg[1 - (\lambda^A+2 \lambda^B) L  \ln \bigg(\frac{M^2_G}{\mu^2}\bigg) \bigg] \delta M_G^2\ = \ 0\times \delta t\ =\ 0\;,
\end{equation}
which implies
\begin{equation}
\delta M_H^2\ =\ \delta M_G^2\ =\ 0\;.
\end{equation}
This means that the solutions to the $H$- and $G$-boson EoMs in the HF approximation are both independent of $\mu$, provided the 2PI RG equations~\eqref{eq:running_HF_B}--\eqref{eq:running_HF_m} are used. Since the SI2PI effective potential is obtained precisely from the solution to the EoM of the Goldstone propagator [cf.~\eqref{eq:SI2PIpotential}], we conclude that \emph{the SI2PI effective potential is exactly RG invariant}, at least in the HF approximation. The same statement holds true for the full one-loop truncated SI2PI potential, which includes the sunset approximation, but the proof is more laborious and is therefore presented in~\ref{app:RG_invariance}. As these findings are highly non-trivial, we feel tempted to conjecture that this basic property of exact RG invariance of the SI2PI potential will persist for any arbitrary high loop-order truncation of the 2PI effective action\footnote{It would be interesting to explore whether this property holds true for the traditional, `unimproved' 2PI effective potential as well. To address this question requires an extensive and independent study, as the EoM for the field~$\phi$ differs from the one that would be obtained in the SI2PI formalism.}.

\begin{figure}
\centering
\includegraphics[height=0.4\textwidth]{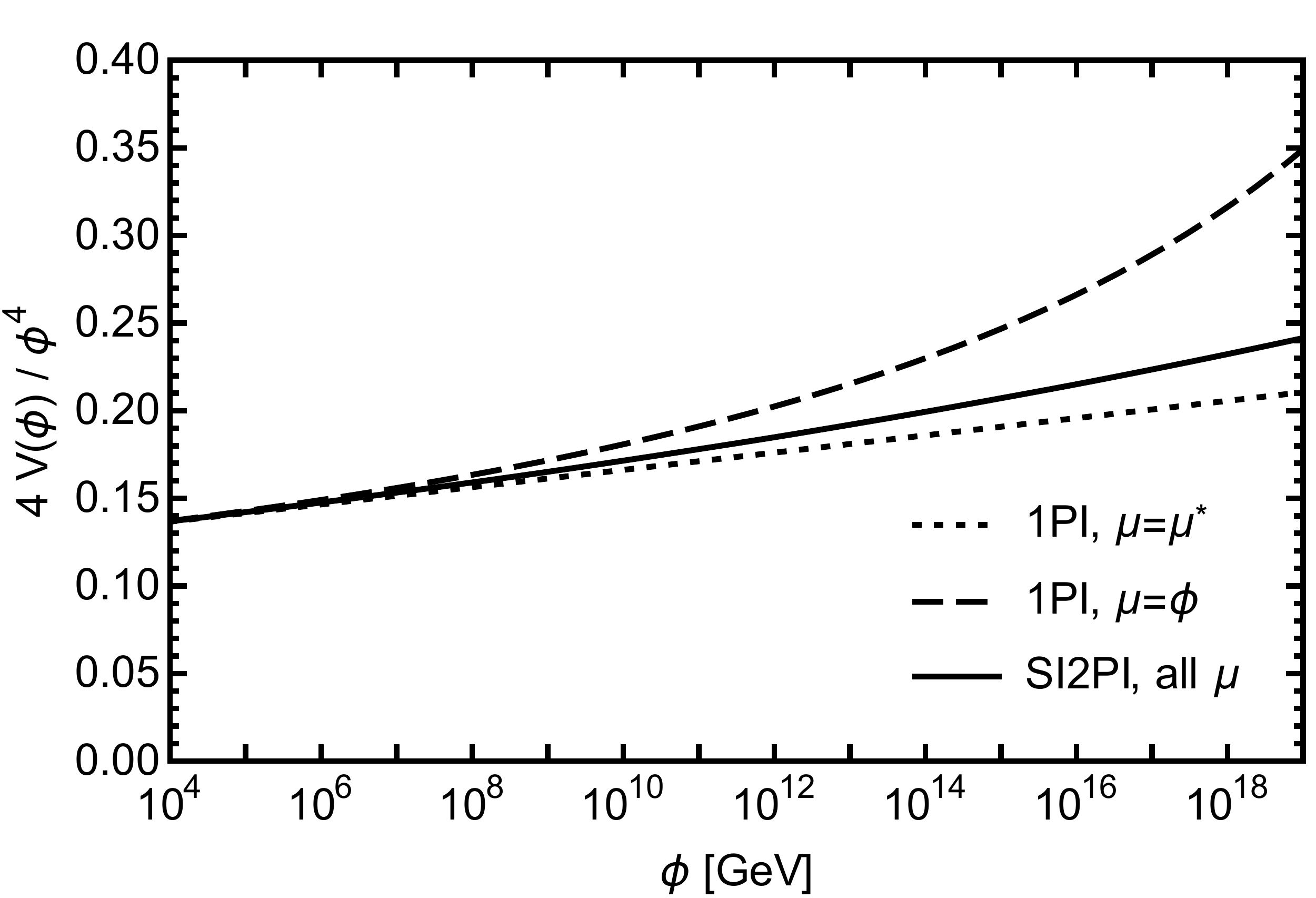}
\caption{The effective potential calculated with different methods. The solid line is the SI2PI one in the sunset approximation. The 1PI effective potential is calculated at the one-loop order, both unimproved (dotted line) and RG-improved (dashed line), with one-loop running. The parameters are as in Fig~\ref{fig:running}. For numerical purposes, we have worked in the massless approximation $m \simeq 0$, since mass effects for field values much larger than the electroweak scale are expected to be negligible. \label{fig:effpot_exactRG}}
\end{figure}

In Fig.~\ref{fig:effpot_exactRG} we plot the SI2PI effective potential in the sunset approximation, along with the 1PI unimproved ($\mu = \mu^*$) and RG-improved ($\mu = \phi$) one-loop effective potentials. Here and in the following, we neglect the electroweak-scale mass $m$, since we are interested in large field values $\phi \gg m$. As expected, we see that the difference between the unimproved and improved 1PI effective potentials becomes significant for high field values $\phi$. Instead, the SI2PI effective potential is $\mu$-independent, yielding the same value for any choice of $\mu$. Nevertheless, we see that the results for the SI2PI effective potential differ significantly from the corresponding ones obtained in the RG-improved 1PI framework. In the next section we identify the origin of this difference, which we address.  As a consequence, this will enable us to develop a novel method to calculate the effective potential semi-perturbatively, thereby increasing considerably its precision in comparison to the standard RG-improved 1PI approach.

\section{Renormalization Group and Symmetry Improved 2PI Effective Potential}\label{sec:RGSI2PI}

In the previous section, we have shown that the complete one-loop truncated SI2PI effective potential is exactly RG invariant, provided all the 2PI parameters run as given by~\eqref{eq:running_sunset_lambda}--\eqref{eq:running_sunset_m} in the sunset approximation. Nevertheless, the predictions obtained in this SI2PI framework for the effective potential differ from those found for the RG-improved 1PI effective potential. In this section, we first clarify the origin of this difference and then present a new computational method of higher precision for the effective potential by unifying the two theoretical approaches of RG improvement and Symmetry improvement in the context of the 2PI effective action. We call the effective potential computed with this new method the Renormalization Group  and Symmetry Improved 2PI (RG-SI2PI)  effective potential.

\begin{figure}
\centering
\includegraphics[height=2.cm]{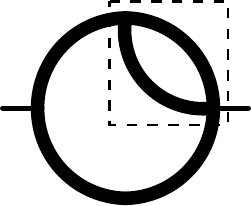} \qquad\qquad
\includegraphics[height=2.cm]{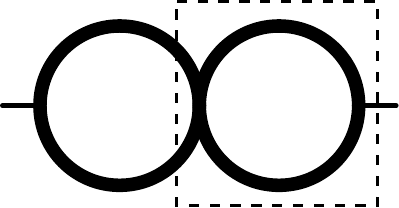} \qquad\qquad
\includegraphics[height=1.8cm]{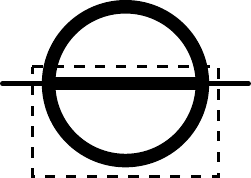}
\caption{The two-loop 2PI self-energies whose subdivergences contribute to the $O(L)$ running. As can be seen by comparison with~\eqref{eq:eombareHscal} and~\eqref{eq:eombareGscal}, the first two diagrams contribute to the running of $\lambda$, and the third diagram to the running of $\lambda_2^{A,B}$. \label{fig:missing_diagrams}}
\end{figure}

To start with, let us first compare the RG equations of the 2PI quartic couplings \eqref{eq:running_sunset_lambda}--\eqref{eq:running_sunset_l2B} with the one for the 1PI 
quartic coupling $\lambda_{\rm 1PI}$ at the one-loop level,
\begin{equation}\label{eq:running_1PI_1loop}
\displaystyle \mu \frac{d}{d \mu} \lambda_{\rm 1PI} \ = \ - 2 \epsilon \, \lambda_{\rm 1PI}  \, + \, 20 L \lambda_{\rm 1PI}^2\;.
\end{equation}
We observe that, if we take all 2PI quartic couplings equal in \eqref{eq:running_sunset_lambda}--\eqref{eq:running_sunset_l2B}, some $O(L)$ contributions are still missing. In fact, the omission of these contributions is exactly the origin of the limited accuracy of the predictions for the  SI2PI effective potential through this order. The remaining $O(L)$ terms are generated when one includes the higher-order 2PI self-energy diagrams as shown in Fig.~\ref{fig:missing_diagrams}. In particular, their $O(L)$ contribution to the RG running of the quartic couplings is described by the extra terms 
\begin{align}
\mu \frac{d}{d \mu} \lambda \ &\supset\ 20 L \lambda^2 \, +\, O(L^2)\;,\\
\mu \frac{d}{d \mu} \lambda_2^A \ &\supset\ 8 L \lambda^2 \, +\, O(L^2)\;,\\
\mu \frac{d}{d \mu} \lambda_2^B \ &\supset\ 16 L \lambda^2 \, +\, O(L^2)\;.
\end{align}
When these terms are added to the RHS of the RG equations~\eqref{eq:running_sunset_lambda}, \eqref{eq:running_sunset_l2A} and \eqref{eq:running_sunset_l2B}, the 1PI $O(L)$ result \eqref{eq:running_1PI_1loop} is recovered in the limit where all the quartic couplings are taken to be equal. In other words, although the contribution of the diagrams in Fig.~\ref{fig:missing_diagrams} to the EoMs is subleading in the sunset approximation, their contribution to the running of the couplings is still at the same leading order. Therefore, without going through the strenuous procedure of 2PI renormalization at higher orders, one practical way to obtain accurate results will be to appropriately amend the $\beta$-functions as given by ordinary perturbation theory through the desired loop order for all the 2PI couplings and calculate the SI2PI effective potential using these improved RG equations. We will adopt this novel method below in order to compute a RG and Symmetry improved 2PI effective potential for a $\mathbb{O}(2)$ theory, beyond the NNLO.

In detail, the  RG equations that we will be utilizing  with this method are
\begin{align}
   \label{eq:RGlambda}
\mu \frac{d}{d \mu} \bar\lambda \ &= \ 20 L \bar\lambda^2 \, -\, 240 L^2 \bar\lambda^3 \,+\, 8 L^3 \big(617+ 384 \zeta(3)\big) \bar\lambda^4 \;,\\
   \label{eq:RGphibar}
\mu \frac{d}{d \mu} \bar\phi \ &= \ - \big( 4 L^2 \lambda^2 - 20 L^3 \lambda^3 \big) \, \bar \phi \; ,
\end{align}
where $\bar\lambda$ collectively denotes all running quartic couplings and $\bar \phi =\bar{\phi} (\mu )$ is the running field, which enters both the RG-SI2PI effective potential and the corresponding NNLO-improved 1PI effective potential.  In~\eqref{eq:RGlambda} and~\eqref{eq:RGphibar}, we use the perturbative three-loop $\beta$- and $\gamma$-functions for the quartic coupling $\lambda$ and the running field $\bar{\phi}$ (see e.g.~\cite{Kleinert:1991rg}). 
By construction, within the adopted method, the so-derived RG-SI2PI effective potential contains all the two-loop graphs present in the respective perturbative 1PI effective potential.  In addition, the RG-SI2PI effective potential contains selective fractal-type resummations of higher-order diagrams as sketched in Fig.~\ref{fig:resum1loop} and~\ref{fig:resum2loop}, whereas the running of the field~$\bar{\phi}$ and of the quartic couplings is the same as that used for evaluating the corresponding 1PI effective potential.

Within the approximative method that we follow here, the RG-SI2PI effective potential is no longer exactly RG invariant, because {\em not all} restoring corrections of RG invariance as dictated by the 2PI formalism were introduced to the running of the kinematic parameters.
Thus, in order to minimize the impact of the residual large logarithms, we may choose a sliding renormalization scale $\mu = O(1) \times \bar \phi$. This is standard procedure for RG improvement of the 1PI effective potential, but it  introduces some subtleties in the SI2PI framework. For a $\phi$-dependent RG scale $\mu =\mu(\phi)$, the 1PI Ward Identity~\eqref{eq:SI2PIpotential}, used to obtain the SI2PI effective potential, gives now only the \emph{partial} derivative of the effective potential:
\begin{equation}\label{eq:part_der}
\frac{\partial}{\partial \bar\phi} V\big(\bar \phi; \bar\lambda(\mu(\bar \phi)),\mu(\bar \phi)\big) \ = \ - \bar \phi \, \Delta_G^{-1}\big(k=0, \bar \phi; \bar\lambda(\mu(\bar \phi)),\mu(\bar \phi)\big) \;,
\end{equation}
where the running field $\bar \phi$ becomes now an implicit function of $\phi$: $\bar \phi(\mu = \mu^*) = \phi$. However, in order to compute the SI2PI effective potential, we are interested in the \emph{total} derivative
\begin{equation}\label{eq:tot_der}
\frac{d}{d \bar\phi} V\big(\bar \phi; \bar \lambda(\mu(\bar \phi)),\mu(\bar \phi)\big) \ = \ \frac{\partial V}{\partial \bar \phi} \;+\; \frac{\partial V}{\partial \bar\lambda} \, \frac{d \bar \lambda}{d \mu} \, \frac{d \mu}{d \bar \phi} \;+\; \frac{\partial V}{\partial \mu} \,  \, \frac{d \mu}{d \bar \phi} \;.
\end{equation}
Hence, this last relation can be rewritten in the equivalent integral form as
\begin{equation}
V(\phi) \ = \ \int_0^{\bar \phi(\phi)} \!\!\! d\phi' \, \frac{\partial V}{\partial \phi'}\big(\phi'; \bar\lambda(\mu),\mu\big) \bigg|_{\mu(\bar \phi(\phi))} \;.
\end{equation}
While the first term on the RHS of~\eqref{eq:tot_der} is obtained by virtue of~\eqref{eq:part_der}, the remaining terms need to be approximated by their 1PI expressions through the two-loop order 
considered here. These additional terms can be computed, given the explicit two-loop order expression
for the 1PI effective potential given in Appendix C~[cf.~\eqref{eq:V1PI}]. In this way, one can integrate \eqref{eq:tot_der} with respect to $\bar \phi$ and obtain $V(\phi) = V(\bar \phi(\phi))$. Notice that this 
additional step is introduced by the procedure of RG-improving the effective potential, 
within our approximative method. Instead, it is not necessary for the exactly RG-invariant SI2PI effective potential calculated in Sec.~\ref{sec:RG_inv}. To make this explicit, note that the latter satisfies
\begin{equation}
V\big(\bar \phi; \bar\lambda(\mu(\bar \phi)),\mu(\bar \phi)\big) \ = \ V\big(\bar \phi; \bar\lambda(\mu=\mu^*),\mu=\mu^*\big) \;,
\end{equation}
so that the total and partial derivatives coincide, which implies that the second and third term on the RHS of~\eqref{eq:tot_der} cancel exactly.

\begin{figure}
\centering
\includegraphics[height=0.4\textwidth]{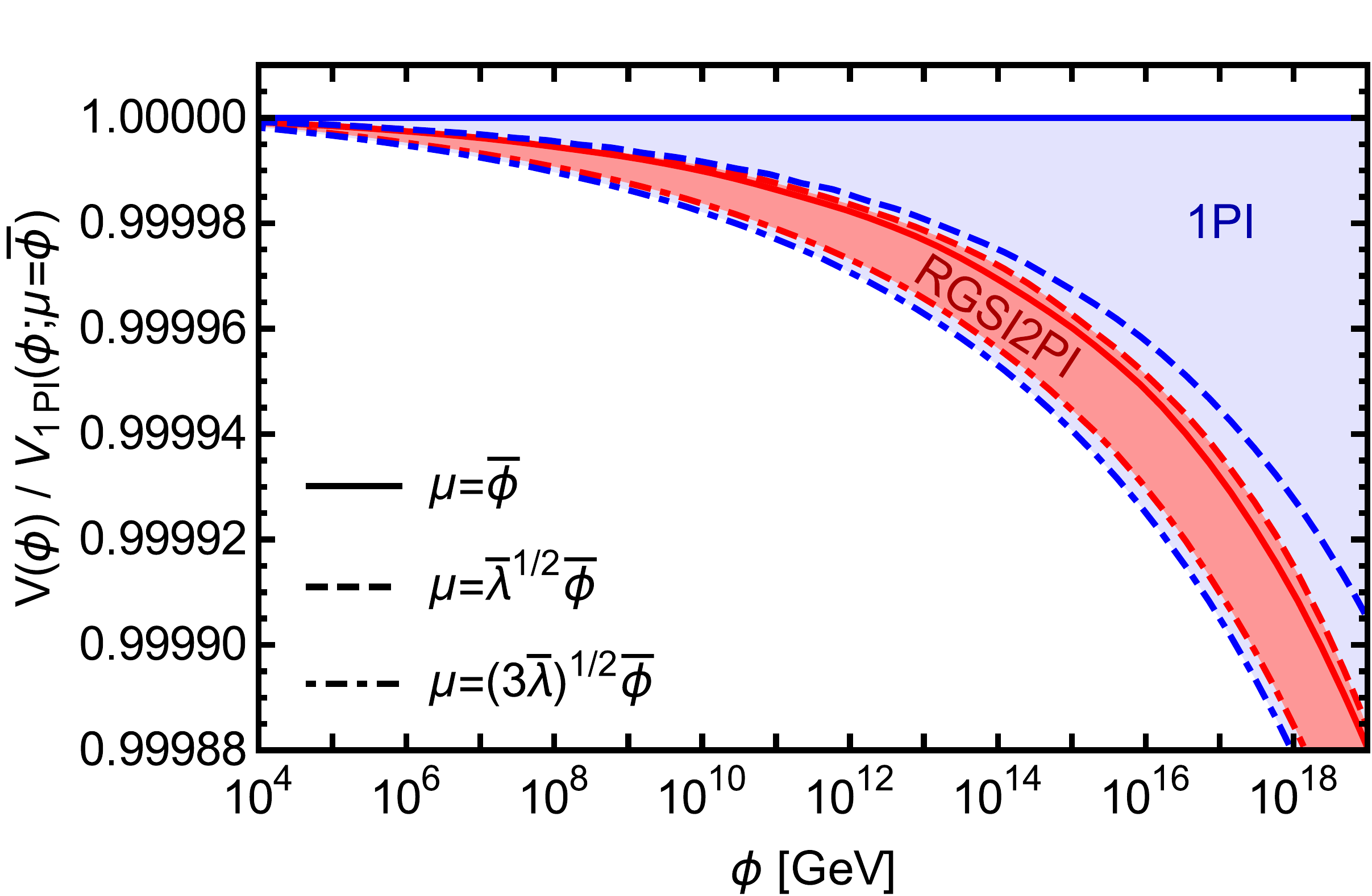}
\caption{The RG-improved effective potential calculated in the two-loop 1PI formalism (blue lines) and the SI2PI one (red lines), for different choices of the renormalization scale $\mu$. We have normalized the results to the 1PI $\mu = \bar \phi$ case, for convenience of presentation. The area between the different blue and red lines provides an estimate for the accuracy of the results obtained in the two methods. The input values at the scale $\mu^* = \unit[125]{GeV}$ are given in Table~\ref{tab:lambda}.\label{fig:RGSI2PI}}
\end{figure}

Following the approximative method outlined above, we can now solve numerically the 2PI EoMs~\eqref{eq:eomren_H} and~\eqref{eq:eomren_G}. We remind the reader that as described in Sec.~\ref{sec:SI2PI}, these include contributions from one- and two-loop 2PI self-energies; the latter were approximated by their perturbative 1PI forms. We have adapted the numerical code  in~\cite{Pilaftsis:2013xna} by including these two-loop 2PI contributions and a $\phi$-dependent renormalization scale $\mu = \mu (\bar \phi)$ (see also discussion below). We first solve numerically the coupled system of EoMs for the $H$- and $G$-propagators for different values of $\bar \phi$, and then utilize the solution for the $G$-propagator in~\eqref{eq:part_der} to find the SI2PI effective potential by means of~\eqref{eq:tot_der}. To maximize the numerical accuracy, we have rescaled the EoMs by re-expressing all mass-dimensional terms in units of $\bar \phi$, so that no large mass hierarchies are present during the numerical computation. Also, in the large-field $\phi$ regime of interest here, we have safely set $m\simeq 0$. We have used grids of 1001 points in momentum space, with a UV cutoff $\Lambda = 50 \, \bar \phi$, tolerating a relative threshold error of $10^{-6}$ for the convergence of the numerical algorithm. The IR cutoff is chosen to be $10^{-5} \, \bar \phi$, and we have checked that all our predictions for the RG-SI2PI effective potential remain stable against variations of either the UV or the IR cutoff. These checks provide firm support of our regularization treatment in numerically solving the relevant 2PI EoMs in the sunset approximation, whilst the potential IR-sensitive effects claimed in~\cite{Marko:2016wtw} have no noticeable impact on the physical quantities that we have been studying.

In Fig.~\ref{fig:RGSI2PI} we present numerical estimates of the RG-SI2PI effective potential
and the corresponding RG-improved 1PI effective potential through NNLO, as functions of the 
field~$\phi$. The predicted values are displayed for several choices of the RG scale $\mu$. The first obvious choice is $\mu = \bar \phi$. In addition, we have considered other relevant mass-scales of the theory, such as  $M_H \simeq \sqrt{3\lambda} \bar \phi$ and $M_G \simeq \sqrt{\lambda} \bar \phi$. Following \cite{Bando:1992np}, we have taken $\mu$ to be equal to either of them, i.e.~$\mu = \sqrt{3\lambda}\, \bar \phi$ and $\mu = \sqrt{\lambda}\, \bar \phi$. For all the different cases, we have chosen the value of $\lambda$ at the scale $\mu^* = \unit[125]{GeV}$, such that the value of the effective potentials calculated at different RG scales coincides at $\phi \approx \mu^*$. The precise values of $\lambda(\mu^*)$ required to achieve to  this are given in Table~\ref{tab:lambda}.

\begin{table}
\centering
\begin{tabular}{c|l}
method & $\lambda(\mu^*)$\\
\hline
\rule{0pt}{13pt}SI2PI $\mu = \sqrt{\lambda}\, \bar \phi$ & 0.13\\
\rule{0pt}{13pt}SI2PI $\mu = \sqrt{3 \lambda}\, \bar \phi$ & 0.13000235\\
\rule{0pt}{13pt}SI2PI $\mu = \bar \phi$ & 0.13000469\\
\rule{0pt}{13pt}1PI $\mu = \sqrt{\lambda}\, \bar \phi$ & 0.13000003\\
\rule{0pt}{13pt}1PI $\mu = \sqrt{3 \lambda}\, \bar \phi$ & 0.13000287\\
\rule{0pt}{13pt}1PI $\mu = \bar \phi$ & 0.13000985\\
\end{tabular}
\caption{The input values for the scalar quartic coupling~$\lambda$ for different RG-improved methods utilized to calculate the effective potential, plotted in Fig.~\ref{fig:RGSI2PI}, chosen so as to match the different lines at the electroweak scale~$\mu = \mu^* = 125$~GeV.
\label{tab:lambda}}
\end{table}

The bands formed by the different lines in Fig.~\ref{fig:RGSI2PI} give a kind of a measure for the $\mu$-dependence of the theoretical predictions, and as such, they should be viewed as theoretical uncertainties inherent to these NNLO calculations. We observe that the bands predicted for the RG-SI2PI potential all lie within the respective ones for the 1PI effective potential, which clearly shows that the predictions for the former potential suffer less from $\mu$-dependent theoretical uncertainties.  Most impressively, the uncertainty of the SI2PI effective potential is significantly smaller than the 1PI one, which may be reduced by up to one order of magnitude when compared with the 1PI results, e.g.~for the bands restricted by the lines $\mu =\bar{\phi}$ and $\mu = \sqrt{3 \lambda}\, \bar \phi$. These findings should not be too surprising, as the RG-SI2PI effective potential, albeit not exactly RG invariant, carries a sort of a {\em memory} effect of exact RG invariance that the SI2PI formalism seems to be endowed with.

\vfill\eject

\section{Conclusions}\label{sec:conclusions}

We have studied the RG properties of the Symmetry Improved Two-Particle-Irreducible effective action truncated to two-loop order, within the context of a simple $\mathbb{O}(2)$ scalar theory. We have been able to prove a number of remarkable field-theoretical results, which to the best of our knowledge have not been discussed adequately in the existing literature.

Unlike the perturbative 1PI effective action, the loop-truncated 2PI effective action of a simple $\phi^4$-theory requires different renormalizations for all mass and coupling parameters, in order to obtain UV finite results for all possible operators consisting of the renormalized field $\phi$ and its dressed propagator $\Delta$. Working in this loop-truncated 2PI framework, we have shown that the RG equations for all relevant 2PI masses and couplings of a simple $\mathbb{O}(2)$ model are {\em exactly} UV finite in both the HF and sunset approximations, as all $1/\epsilon^n$ poles of order $n = 2$ and higher vanish identically in dimensional regularization.  This is the first important result of our study, which is highly non-trivial.  In fact, in ordinary 1PI perturbation theory, such cancellations of the higher order $1/\epsilon$ poles occur at different loop orders, and they become only complete to all orders.
As shown in Fig.~\ref{fig:running}, the different 2PI couplings exhibit different RG runnings, even though they are taken to be equal at some given scale $\mu = \mu^*$. Note that this variance in the running of the quartic couplings is expected to decrease, as the order of loop truncation of the 2PI effective action increases. Thus, we expect to obtain a universal RG running for all 2PI quartic couplings to all loop orders.

Another important result of our study was to prove that the one-loop truncated SI2PI effective potential is \emph{exactly} RG invariant in the HF and the sunset approximations. This is an extraordinary property and should be contrasted with what happens in the usual 1PI formalism, where the effective potential becomes RG invariant, only upon ignoring {\em unknown} higher order terms. Therefore, whilst one needs to do educated guesses for the RG scale $\mu$ so as to RG-improve the effective potential and minimize the impact of these unknown terms in the 1PI formalism, there is no need to do so in the 2PI framework, as the SI2PI effective potential is {\em exactly} $\mu$-independent and as such, its value will be equal for both fixed $\mu = \mu^*$ and sliding $\mu = \mu (\phi )$, e.g.~$\mu = \phi$.

However, in the sunset approximation that we have been considering here, not all RG equations of the 2PI couplings have the required accuracy, as not all of them match to the one-loop $\beta$-function, $\beta_\lambda$, known from the standard 1PI perturbation theory. We have identified the origin of the missing terms in the one-loop 2PI $\beta$-functions, which stem from higher order loop-graph topologies in the 2PI effective action that first appear at the three-loop order of truncation. To improve the accuracy of the RG running, we have adopted an approximative method, where the three-loop order 1PI $\beta$- and $\gamma$-functions were utilized for all 2PI quartic couplings and the running field $\phi$. In this approximative method, the exact RG invariance of the SI2PI effective potential is lost,  
so we have resorted to  a RG-improvement approach similar to the one performed for the usual 1PI effective potential. The so-derived RG-SI2PI effective potential contains all two-loop diagrams present in the 1PI effective potential, including the three-loop RG-running of the quartic coupling and the radial scalar field $\phi$. In addition, the RG-SI2PI effective potential goes beyond the NNLO, as it contains selective resummations of infinite series of graphs as illustrated in Figs.~\ref{fig:resum1loop} and~\ref{fig:resum2loop}, which are inherent in the 2PI formalism.

As shown in Fig.~\ref{fig:RGSI2PI}, the RG-SI2PI effective potential gives rise to predictions
that suffer less from $\mu$-dependent theoretical uncertainties. Specifically, we have found that the $\mu$-dependent uncertainties of the SI2PI effective potential are considerably smaller than the one obtained in the 1PI formalism. In particular, the theoretical uncertainties may be lowered by up to one order of magnitude when compared to those obtained in the 1PI framework, as can be seen from  the bands confined by the lines $\mu = \bar{\phi}$ and $\mu = \sqrt{3 \lambda}\, \bar \phi$ in Fig.~\ref{fig:RGSI2PI}. This last finding of our study underlies the remarkable RG properties of the RG-SI2PI effective potential. On the other hand, this should not be considered too surprising, as the RG-SI2PI effective potential seems to have inherited a good degree of exact RG invariance from the original SI2PI formalism.

The present study opens up new exciting vistas that deserve further detailed and rigorous exploration. In particular, it would be interesting to extend the present computation of the SI2PI effective potential by including dressed chiral fermion propagators, and check its invariance under RG running in both the HF and sunset approximations.  This would be an important milestone in this line of investigations, enabling for the first time higher order precision computations for the SM effective potential beyond the NNLO with considerably reduced theoretical uncertainties. An equally interesting avenue that one could follow would be to go beyond the sunset approximation, and investigate whether the exactness of RG invariance persists for the SI2PI effective potential.  The ultimate goal of these studies would be to obtain an {\em exactly} RG-invariant SI2PI effective potential beyond the NNLO, which in turn will help us to obtain more accurate determinations of several physical processes, such as vacuum-to-vacuum tunnelling transitions, beyond their state of the art.

\section*{Acknowledgements}
\vspace{-3mm}
\noindent
The work of A.P. is supported by the Lancaster--Manchester--Sheffield Consortium for Fundamental Physics, under STFC research grant: ST/L000520/1. The work of D.T. is funded by a ULB Postdoctoral Fellowship.

\newpage

\appendix

\makeatletter
\gdef\thesection{\appendixname~\@Alph\c@section}%
\makeatother

\section{Renormalization in the Sunset Approximation}\label{app:sunset}

The renormalization in the sunset approximation is described in detail in~\cite{Pilaftsis:2013xna}. Here we present useful formulae for the renormalized tadpole and sunset integrals, $\mathcal{T}_{a}^{\rm fin}$ and $\mathcal{I}_{ab}^{\rm fin}$, and for the 2PI quartic coupling CTs $\delta\lambda^{A,B}_1$ and $\delta\lambda^{A,B}_2$ in~DR, which are relevant to our study. Here we assume that all the renormalized 
proper 2PI couplings, $\lambda^{A,B}_1$ and $\lambda^{A,B}_2$, are, in principle, different from each 
other. In this way, we generalize the results of~\cite{Pilaftsis:2013xna}, beyond the HF approximation.

Working in the Euclidean momentum space, we may conveniently express the renormalized sunset integral as 
\begin{equation}\label{eq:ren_I}
\mathcal{I}_{ab}^{\rm fin}(p) \ \equiv \  \int_k \bigg( \Delta_a(k-p)\, \Delta_b(k) \:-\: \Delta_0(k)^2\bigg) \;,
\end{equation}
where $a,b = H,G$, and
\begin{equation}
\Delta_0(k)\ \equiv \ \big(k^2 + \mu^2\big)^{-1} 
\end{equation}
is an auxiliary propagator introduced here with the purpose to remove UV infinities in~DR.

In order to calculate the renormalized tadpole integral, we parametrize the renormalized propagators as \begin{equation}
    \label{eq:prop_param} 
\Delta_a(k)^{-1} \ = \ k^2 \,+\, M_a^2 \,+\, \Sigma_a(k) \;, 
\end{equation} 
where $M_a^2$ stands for all momentum-independent contributions that originate from the tree-level terms and the tadpole integrals in the renormalized EoMs~\eqref{eq:eomren_H} and~\eqref{eq:eomren_G}, whereas $\Sigma_a(k)$ is the contribution of the sunset integrals, i.e.  
\begin{align}
\Sigma_H(k) \ &\equiv \ - \, 18 \lambda^2 \phi^2 \, \hbar \mathcal{I}_{HH}^{\rm fin}(k) \; - \; 2 \lambda^2 \phi^2 \, \hbar \mathcal{I}_{GG}^{\rm fin}(k) \;, \label{eq:sigmaH}\\
\Sigma_G(k) \ &\equiv \ - \, 4\lambda^2 \phi^2 \, \hbar \mathcal{I}_{HG}^{\rm fin}(k) \;. \label{eq:sigmaG} \end{align} 
The renormalized tadpole integral is then found to be 
\begin{align}
\mathcal{T}_a^{\rm fin} \ &\equiv \ \int_k \Big[ \Delta_a(k) \:-\: \Delta_0(k)  \: - \: \Delta_0(k)^2 
\Big( \mu^2 - M^2_a + \nu_a \, \lambda^2 \phi^2 \hbar\mathcal{I}_{00}^{fin}(k)\Big)  \Big] \notag\\
                & \quad - \: \frac{\mu^2}{16 \pi^2} \: + \: \hbar \, \frac{\nu_a \, \lambda^2 \phi^2}{(16 \pi^2)^2} \,\frac{\eta}{2} \;, \label{eq:ren_T} 
\end{align} 
where $\nu_H = 20$, $\nu_G = 4$ and 
\begin{equation} \eta \ = \ 1 \; - \; \frac{4 i}{\sqrt{3}} \left( \mathrm{Li}_2\frac{1 - i \sqrt{3}}{2} - \frac{\pi^2}{36}\right) \ \simeq \ -1.34391\;.  
\end{equation}

Requiring now that the unrenormalized EoMs~\eqref{eq:eombareHscal} and~\eqref{eq:eombareGscal}
be UV finite, we may derive the 2PI CTs~$\delta\lambda^A_1$ and $\delta\lambda^B_1$ in the 
sunset approximation,
\begin{align}
\delta \lambda_1^A \ &= \ \frac{1}{\displaystyle \bigg( 1 - 2 \frac{L}{\epsilon} \lambda_2^B\bigg)\bigg( 1 - 2 \frac{L}{\epsilon} (\lambda_2^A+\lambda_2^B)\bigg)} \notag\\
&\times \ \bigg\{ 2 \, \frac{L}{\epsilon} \,  \bigg[ \lambda_1^A \lambda_2^A + \lambda_1^B \lambda_2^A + \lambda_1^A \lambda_2^B + 2 \lambda^2  \bigg]  \notag\\
&- \; 4 \, L^2 \bigg[ \frac{1}{\epsilon} (3 \lambda_2^A \lambda^2 +  \lambda_2^B \lambda^2 )
 + \frac{1}{\epsilon^2} (\lambda_1^A \lambda_2^A \lambda_2^B + \lambda_1^A {\lambda_2^B}^2 - \lambda_2^A \lambda^2 + 3 \lambda_2^B \lambda^2) \bigg] \notag\\
 &+ 8 \, L^3 \bigg[ \frac{1}{\epsilon^2} (\lambda_2^A  \lambda_2^B \lambda^2 + {\lambda_2^B}^2 \lambda^2 ) + \frac{1}{\epsilon^3} (\lambda_2^A \lambda_2^B \lambda^2 + {\lambda_2^B}^2 \lambda^2 ) \bigg] \bigg\} \;, \displaybreak[1]\\[6pt]
 \delta \lambda_1^B \ &= \ \frac{1}{\displaystyle 1 - 2 \frac{L}{\epsilon} \lambda_2^B} \bigg\{ 2 \frac{L}{\epsilon} (\lambda_1^B \lambda_2^B + 4 \lambda^2) \,-\, 8 \, L^2 \bigg( \frac{1}{\epsilon} + \frac{1}{\epsilon^2}\bigg) \lambda_2^B \lambda^2 \bigg\} \;.
\end{align}
Instead, the 2PI CTs $\delta m^2$, $\delta \lambda_2^A$ and $\delta \lambda_2^B$ may be obtained by \eqref{eq:dl_A}, \eqref{eq:dl_B} and \eqref{eq:dm}, upon replacing $\lambda^{A,B}$ with $\lambda_{2}^{A,B}$.

\section{Exact RG Invariance in the Sunset Approximation}\label{app:RG_invariance}

In this appendix we prove that the 2PI EoMs~\eqref{eq:eomren_H} and~\eqref{eq:eomren_G} are {\em exactly} RG invariant in the sunset approximation, in close analogy to the proof presented in Sec.~\ref{sec:RG_inv} in the HF approximation.

To start with, let us first point out that in the sunset approximation, it is important to also include the RG runnings ${\cal O}(\epsilon)$ for the field~$\phi$ and the quartic coupling~$\lambda$
that are induced at the tree level in~DR.
These tree-level effects ${\cal O} (\epsilon)$  give non-trivial finite contributions when multiplied with UV poles ${\cal O} (1/\epsilon)$. Thus, under an infinitesimal RG shift: $t \to t + \delta t$, with $t \equiv 
\ln \mu$, the 2PI kinematic parameters transform as follows:
\begin{align}
\delta \phi^2  &= 2 \epsilon \phi^2 \delta t \;,& \delta m^2 &= m^2 \gamma_{m^2} \delta t \;, \notag \\
\delta \lambda &= - 2 \epsilon \lambda \delta t \;, & \delta \lambda_X^Y &= - 2 \epsilon \lambda_X^Y + \beta_X^Y \delta t\;,
\end{align}
where $X = 1,2$ and $Y = A,B$, and the $\beta$- and $\gamma$-functions can be read off from \eqref{eq:running_sunset_lambda}--\eqref{eq:running_sunset_m}.

Our next step is to  calculate the variation of the renormalized sunset and tadpole integrals stated in \eqref{eq:ren_I} and \eqref{eq:ren_T}, respectively, under the same infinitesimal shift: $t \to t + \delta t$. To do so, we  first consider the integral~$\mathcal{I}^{\rm fin}_{ab}$ in~\eqref{eq:ren_I}. In this integral, the term depending on the auxiliary propagator~$\Delta_0(k)$ can be integrated out analytically in~DR. 
The latter allows us to write the unrenormalized sunset integral $\bm{\mathcal I}_{ab}(p)$ as follows:
\begin{equation}
\hbar \bm{\mathcal I}_{ab}(p) \ \equiv \ \hbar \, \overline{\mu}^{2 \epsilon} \!\! \int_k \Delta_a(k) \Delta_b(p-k) \ = \ \hbar \mathcal I_{ab}^{\rm fin}(p)  \;+\; \frac{L}{\epsilon} \;.
\end{equation}
Hence, the variation of $\mathcal{I}^{\rm fin}_{ab}$ is given by
\begin{align}\label{eq:var_I}
\delta (\mathcal{I}^{\rm fin}_{ab}) \ &= \ 2 \epsilon \, \hbar \bm{\mathcal I}_{ab} \delta t \, + \, \hbar \overline{\mu}^{2 \epsilon} \!\! \int_k \delta \big[\Delta_a(k) \Delta_b(p-k) \big] \ = \ 2 L \delta t \,+\,  2 \epsilon \, \hbar \mathcal I_{ab}^{\rm fin} \,+\, {\cal O}(\delta \Delta) \notag\\
& = \ 2 L \delta t \,+\, {\cal O}(\epsilon) \,+\, {\cal O}(\delta \Delta)\;,
\end{align}
where only the leading terms in a series expansion in powers of~$\epsilon$ and~$\delta \Delta \equiv \delta \Delta_{H,G}$ are kept.

Proceeding analogously for the tadpole integral~\eqref{eq:ren_T}, we get
\begin{equation}
\hbar \bm{\mathcal T}_{\!\!a} \ \equiv \ \hbar \, \overline{\mu}^{2 \epsilon} \!\! \int_k \Delta_a(k)\ = \ \hbar \mathcal{T}_a^{\rm fin} \,-\, M^2_a \frac{L}{\epsilon} \, +\, \nu_a \lambda^2 \phi^2 \frac{L^2}{2} \bigg( \frac{1}{\epsilon} - \frac{1}{\epsilon^2}\bigg) \;.
\end{equation}
Given that $\delta(\lambda^2 \phi^2) = - 2 \epsilon \lambda^2 \phi^2 \delta t$, we have
\begin{align}\label{eq:var_T}
\delta(\hbar \mathcal{T}_a^{\rm fin}) \ &= \ 2 \epsilon \, \hbar \bm{\mathcal T}_{\!\!a}  \delta t \,+\, {\cal O}(\delta \Delta) \, +\, \frac{L}{\epsilon} \delta M_a^2 \, +\, 2 \epsilon \, \nu_a \lambda^2 \phi^2 \frac{L^2}{2}  \bigg( \frac{1}{\epsilon} - \frac{1}{\epsilon^2}\bigg) \delta t \notag\\
&= \ - 2 L M_a^2 \delta t \,+\, \frac{L}{\epsilon} \delta M_a^2  \, +\, 2 \epsilon \, \nu_a \lambda^2 \phi^2 L^2  \bigg( \frac{1}{\epsilon} - \frac{1}{\epsilon^2}\bigg) \delta t \, + \, {\cal O}(\epsilon) \,+\, 
{\cal O}(\delta \Delta) \;.
\end{align}
As can be seen from the last expression, one key ingredient is the variation $\delta M_a^2$. This can be found by observing that \eqref{eq:prop_param} implies $\delta (M_a^2 + \Sigma_a) = {\cal O}(\delta \Delta)$, with $a = H,G$, and by varying \eqref{eq:sigmaH} and \eqref{eq:sigmaG}. In this way, we have
\begin{equation}
   \label{eq:dM2a}
\delta M_a^2\ =\ -\,\delta \Sigma_a\, +\, {\cal O}(\delta \Delta)\ = \ \nu_a \lambda^2 \phi^2 2 L \delta t \,  + \, {\cal O}(\delta \Delta) \;.
\end{equation}
Notice that we do {\em not} neglect relevant terms ${\cal O}(\epsilon)$, as they can potentially give finite
contributions when multiplied by $1/\epsilon$ poles. Taking such terms into account,
we end up with
\begin{equation}
\delta(\hbar \mathcal{T}_a^{\rm fin}) \ = \ - 2 L M_a^2 \delta t \, +\, 2 L^2 \nu_a \lambda^2 \phi^2 \delta t \, + \, {\cal O}(\epsilon) \, + \, {\cal O}(\delta \Delta) \;.
\end{equation}
Thus, after RG-varying the EoM~\eqref{eq:eomren_H} pertinent to the inverse Higgs propagator~$\Delta_H^{-1}(k)$, we obtain in the limit $\epsilon \to 0$,
\begin{align}
\delta \Delta_H^{-1} \ = \ \delta t \, \Big[ &(\beta_1^A + 2 \beta_1^B) \phi^2 \,-\, m^2 \gamma_{m^2} \notag\\
& +\, (\beta_2^A + 2 \beta_2^B) \hbar \mathcal{T}_H^{\rm fin} \,+\, (\lambda_2^A + 2 \lambda_2^B) (- 2 L M_H^2 \, +\, 2 L^2 \nu_H \lambda^2 \phi^2 ) \notag\\
& +\, \beta_2^A \hbar \mathcal{T}_G^{\rm fin} \,+\, \lambda_2^A (- 2 L M_G^2 \, +\, 2 L^2 \nu_G \lambda^2 \phi^2 ) \, -\, \nu_H \lambda^2 \phi^2 2 L \Big] \, + \, {\cal O}(\delta \Delta) \;.
\end{align}

Equipped now with all key equations for the variations of  $M^2_H$ and $M^2_G$ in~\eqref{eq:dM2a}, the EoMs~\eqref{eq:eomren_H} and~\eqref{eq:eomren_G}, and the $\beta$- and $\gamma$-functions as deduced from~\eqref{eq:running_sunset_lambda}--\eqref{eq:running_sunset_m}, and after some lengthy algebra, we finally obtain
\begin{equation}
   \label{eq:RGexact}
\delta \Delta_H^{-1} \ = \  {\cal O}(\delta \Delta) \, +\, 0\times \delta t\;.
\end{equation}
We note that an analogous result is obtained for $\Delta_G$ as well. Employing the identity of variations: $\delta \Delta_a = -\Delta_a\, \delta \Delta_a^{-1}\, \Delta_a$ on the RHS of~\eqref{eq:RGexact} for $\Delta_H$ and its corresponding counterpart for $\Delta_G$, we get a linear system of two equations:
\begin{align}
   \label{eq:variationA}
A_H\,  \delta \Delta_H^{-1}\: +\: A_G\, \delta \Delta_G^{-1}\ &=\ 0\;,\\
   \label{eq:variationB}
B_H\,  \delta \Delta_H^{-1}\: +\: B_G\, \delta \Delta_G^{-1}\ &=\ 0\; ,
\end{align}
where $A_{H,G}$ and $B_{H,G}$ are lengthy expression depending on $\Delta_{H,G} (k)$, and all other kinematic parameters of the theory, such as the field $\phi$, the mass parameter $m^2$ and the 2PI quartic couplings.
It can be shown that the two equations in~\eqref{eq:variationA} and \eqref{eq:variationB} are linearly independent. This implies that barring fine-tunings,
\begin{equation}
\delta \Delta^{-1}_H\ =\ \delta \Delta^{-1}_G\ =\ 0
\end{equation}
is a non-trivial solution to~\eqref{eq:variationA} and \eqref{eq:variationB}, for arbitrary
values of $k$ and $\phi$. Hence,  the solutions to the 2PI EoMs for the $H$- and $G$- propagators are exactly RG invariant. Therefore, we conclude that the SI2PI effective potential, obtained from $\Delta^{-1}_G(k=0)$, is {\em exactly} RG invariant also in the sunset approximation.

\section{Perturbative Two-Loop Formulae}\label{app:pert_2loop}

In this appendix, we present analytical expressions for the two-loop 1PI effective potential $V_{\rm 1PI}(\phi)$ and the renormalized two-loop self-energies $\Pi^{\mathrm{2PI}, (2)}_{H,G}$  appearing in the EoMs ~\eqref{eq:eomren_H} and~\eqref{eq:eomren_G}, all in the limit of interest $m^2 \to 0$. These are calculated in perturbation theory in the so-called $\overline{\rm MS}$ scheme by means of standard techniques~\cite{Martin:2003it, Martin:2003qz}, and approximated by their zero-momentum value. We adopt the compact notation of~\cite{Martin:2003it}, and introduce  the abbreviations: $\overline{\ln} x \equiv \ln(x/\mu^2)$ and $s=-k^2$, where $k$ is the Euclidean momentum. Moreover, we define the one-loop function
\begin{equation}
A(x) \ \equiv \ x \, (\overline{\ln} x - 1) \;,
\end{equation}
and the two-loop function $I(x,y,z)$ evaluated at zero momentum, whose analytical expression may be found in~\cite{Martin:2003qz}.

The 1PI effective potential for the scalar $\mathbb{O}(2)$ model, up to two-loop order, is given by
\begin{align}\label{eq:V1PI}
V_{\rm 1PI}(\phi) \ &= \ \frac{\lambda \phi^4}{4} \; + \; L \, \frac{\lambda^2 \phi^4}{4} \bigg[ 10 \bigg( \overline \ln (\lambda \phi^2) - \frac{3}{2}\bigg) \, +\, 9 \ln 3\bigg] \notag\\
& + \ L^2 \, \frac{\lambda}{4} \, \Big[ 3 A(h)^2 + 2 A(h) A(g) + 3 A(g)^2 - 12 \lambda \phi^2 I(h,h,h) - 4 \lambda \phi^2 I(h,g,g)\Big]\;,
\end{align}
with $h = 3 \lambda \phi^2$ and $g = \lambda \phi^2$.

The 2PI two-loop topologies contributing to $\Pi^{\mathrm{2PI}, (2)}_{H,G}$ are shown in Fig.~6 of~\cite{Pilaftsis:2015bbs}, after neglecting in that work the fermion and charged Goldstone-boson contributions. For the $\mathbb{O}(2)$ model at hand, we~find
\begin{align}
\Pi^{\mathrm{2PI}, (2)}_H/L^2 \ &= \ 6 \lambda^3 \phi^2 \big[ 9 \, (\overline \ln \, h)^2 + 2 \, (\overline \ln \, h)(\overline \ln \, g) + (\overline \ln \, g)^2\big] \notag \\
&- \ 6 \lambda^2 I(h,h,h) \;-\; 2 \lambda^2 I(h,g,g) \notag\\
&- \ 8 \lambda^3 \phi^2 \big[ 27 \, I(h',h,h) \; +\; 3\, I(h',g,g) \; + \; 2 \, I(g',g,h)\big] \notag \\
&- 8 \lambda^4 \phi^4 \big[ 81 \, I(h',h',h) \; + \; 6 \, I(h',g',g) \;+\; I(g',g',h)\big] \;,\\[6pt]
\Pi^{\mathrm{2PI}, (2)}_G/L^2 \ &= \ 8 \lambda^3 \phi^2 \big[A(h) - A(g) \big]^2/(h-g)^2 \notag \\
&- \ 2 \lambda^2 I(g,h,h) \; - \; 6 \lambda^2 I(g,g,g) \notag \\
&+ \ 8 \lambda^3 \phi^2 \big[ 3 \,I(g,g,g) \;-\; 3 \,I(h,h,h) \;+\; I(g,h,h) \;-\; I(g,g,h)\big]/(h-g) \notag\\
&+ \ 16 \lambda^4 \phi^4 \big[ - I(g,g,g) \,-\,  3 \, I(h,h,h) \, + \, 5 \, I(g,g,h) \,-\,I(g,g,h) \big]/(h-g)^2 \;.
\end{align}
where now $h = (\lambda_1^A + 2 \lambda_1^B) \phi^2$, $g = \lambda_1^A \phi^2$ and the prime ($'$) denotes differentiation of the function $I(x,y,z)$ with respect to the primed argument, 
e.g.~$I(x',y,z) \equiv \partial I(x,y,z)/\partial x$, $I(x,y',z) \equiv \partial I(x,y,z)/\partial y$ etc.


\newpage

\begin{thebibliography}{99}

\bibitem{Bezrukov:2012sa}
  F.~Bezrukov, M.~Y.~Kalmykov, B.~A.~Kniehl and M.~Shaposhnikov,
  JHEP {\bf 1210} (2012) 140
  [arXiv:1205.2893 [hep-ph]].
  
\bibitem{Degrassi:2012ry}
  G.~Degrassi, S.~Di Vita, J.~Elias-Miro, J.~R.~Espinosa, G.~F.~Giudice, G.~Isidori and A.~Strumia,
  JHEP {\bf 1208} (2012) 098
  [arXiv:1205.6497 [hep-ph]].
  
\bibitem{Buttazzo:2013uya}
  D.~Buttazzo, G.~Degrassi, P.~P.~Giardino, G.~F.~Giudice, F.~Sala, A.~Salvio and A.~Strumia,
  JHEP {\bf 1312} (2013) 089
  [arXiv:1307.3536 [hep-ph]].

\bibitem{Bezrukov:2014ina}
  F.~Bezrukov and M.~Shaposhnikov,
  J.\ Exp.\ Theor.\ Phys.\  {\bf 120} (2015) 335
   [Zh.\ Eksp.\ Teor.\ Fiz.\  {\bf 147} (2015) 389]
  [arXiv:1411.1923 [hep-ph]].

\bibitem{Pilaftsis:2013xna}
  A.~Pilaftsis and D.~Teresi,
  Nucl.\ Phys.\ B {\bf 874} (2013) no.2,  594
  [arXiv:1305.3221 [hep-ph]].

\bibitem{Garbrecht:2015cla}
  B.~Garbrecht and P.~Millington,
  Nucl.\ Phys.\ B {\bf 906} (2016) 105
  [arXiv:1509.07847 [hep-th]].
  
\bibitem{Ellis:2015xwp}
  J.~Ellis, N.~E.~Mavromatos and D.~P.~Skliros,
  Nucl.\ Phys.\ B {\bf 909} (2016) 840
  [arXiv:1512.02604 [hep-th]].

\bibitem{Martin:2014bca}
  S.~P.~Martin,
  Phys.\ Rev.\ D {\bf 90} (2014) no.1,  016013
  [arXiv:1406.2355 [hep-ph]].

\bibitem{Elias-Miro:2014pca}
  J.~Elias-Miro, J.~R.~Espinosa and T.~Konstandin,
  JHEP {\bf 1408} (2014) 034
  [arXiv:1406.2652 [hep-ph]].

\bibitem{Pilaftsis:2015bbs}
  A.~Pilaftsis and D.~Teresi,
  Nucl.\ Phys.\ B {\bf 906} (2016) 381
  [arXiv:1511.05347 [hep-ph]].

\bibitem{Garbrecht:2015oea}
  B.~Garbrecht and P.~Millington,
  Phys.\ Rev.\ D {\bf 91} (2015) 105021
  [arXiv:1501.07466 [hep-th]].
  
\bibitem{Garbrecht:2015yza}
  B.~Garbrecht and P.~Millington,
  Phys.\ Rev.\ D {\bf 92} (2015) 125022
  [arXiv:1509.08480 [hep-ph]]. 

\bibitem{Plascencia:2015pga} 
  A.~D.~Plascencia and C.~Tamarit,
  JHEP {\bf 1610}, 099 (2016)
  [arXiv:1510.07613 [hep-ph]].

\bibitem{Espinosa:2016uaw}
  J.~R.~Espinosa, M.~Garny and T.~Konstandin,
  Phys.\ Rev.\ D {\bf 94} (2016) no.5,  055026
  [arXiv:1607.08432 [hep-ph]].

\bibitem{Espinosa:2016nld}
  J.~R.~Espinosa, M.~Garny, T.~Konstandin and A.~Riotto,
  arXiv:1608.06765 [hep-ph].

\bibitem{Branchina:2013jra}
  V.~Branchina and E.~Messina,
  Phys.\ Rev.\ Lett.\  {\bf 111} (2013) 241801
  [arXiv:1307.5193 [hep-ph]].
  
\bibitem{Branchina:2014rva}
  V.~Branchina, E.~Messina and M.~Sher,
  Phys.\ Rev.\ D {\bf 91} (2015) 013003
  [arXiv:1408.5302 [hep-ph]].

\bibitem{Cornwall:1974vz}
  J.~M.~Cornwall, R.~Jackiw and E.~Tomboulis,
  Phys.\ Rev.\ D {\bf 10} (1974) 2428.

\bibitem{Petropoulos:1998gt}
  N.~Petropoulos,
  J.\ Phys.\ G {\bf 25} (1999) 2225
  [hep-ph/9807331].


\bibitem{Pilaftsis:2015cka}
  A.~Pilaftsis and D.~Teresi,
  J.\ Phys.\ Conf.\ Ser.\  {\bf 631} (2015) no.1,  012008
  [arXiv:1502.07986 [hep-ph]].
  
\bibitem{Mao:2013gva}
  H.~Mao,
  Nucl.\ Phys.\ A {\bf 925} (2014) 185
  [arXiv:1305.4329 [hep-ph]].
  
\bibitem{Brown:2015xma}
  M.~J.~Brown and I.~B.~Whittingham,
  Phys.\ Rev.\ D {\bf 91} (2015) no.8,  085020
  [arXiv:1502.03640 [hep-th]].
  
  
\bibitem{Brown:2016sak}
  M.~J.~Brown, I.~B.~Whittingham and D.~S.~Kosov,
  Phys.\ Rev.\ D {\bf 93} (2016) no.10,  105018
  [arXiv:1603.03425 [hep-th]].
  
\bibitem{Marko:2016wtw}
  G.~Mark\'o, U.~Reinosa and Z.~Sz\'ep,
  Nucl.\ Phys.\ B {\bf 913} (2016) 405
  [arXiv:1604.04193 [hep-ph]].

\bibitem{Brown:2016vaj}
  M.~J.~Brown and I.~B.~Whittingham,
  Phys.\ Rev.\ D {\bf 95} (2017) no.2,  025018
  [arXiv:1611.05226 [hep-th]].
  
\bibitem{Carrington:2014lba}
  M.~E.~Carrington, W.~J.~Fu, D.~Pickering and J.~W.~Pulver,
  Phys.\ Rev.\ D {\bf 91} (2015) no.2,  025003
  [arXiv:1404.0710 [hep-ph]].
  
\bibitem{Pawlowski:2015mlf}
  J.~M.~Pawlowski, M.~M.~Scherer, R.~Schmidt and S.~J.~Wetzel,
  arXiv:1512.03598 [hep-th].
  
\bibitem{vanHees:2001ik}
  H.~van Hees and J.~Knoll,
  Phys.\ Rev.\ D {\bf 65} (2002) 025010
  [hep-ph/0107200].
  
\bibitem{Blaizot:2003an}
  J.~P.~Blaizot, E.~Iancu and U.~Reinosa,
  Nucl.\ Phys.\ A {\bf 736} (2004) 149
  [hep-ph/0312085].

\bibitem{Berges:2005hc}
  J.~Berges, S.~Borsanyi, U.~Reinosa and J.~Serreau,
  Annals Phys.\  {\bf 320} (2005) 344
  [hep-ph/0503240].
  
\bibitem{Fejos:2007ec}
  G.~Fejos, A.~Patkos and Z.~Szep,
  Nucl.\ Phys.\ A {\bf 803} (2008) 115
  [arXiv:0711.2933 [hep-ph]].  
  


\bibitem{Machacek:1983tz}
  M.~E.~Machacek and M.~T.~Vaughn,
  Nucl.\ Phys.\ B {\bf 222} (1983) 83.

\bibitem{Kleinert:1991rg}
  H.~Kleinert, J.~Neu, V.~Schulte-Frohlinde, K.~G.~Chetyrkin and S.~A.~Larin,
  Phys.\ Lett.\ B {\bf 272} (1991) 39
   Erratum: [Phys.\ Lett.\ B {\bf 319} (1993) 545],
  [hep-th/9503230].

\bibitem{Bando:1992np}
  M.~Bando, T.~Kugo, N.~Maekawa and H.~Nakano,
  Phys.\ Lett.\ B {\bf 301} (1993) 83
  [hep-ph/9210228].
  
\bibitem{Martin:2003it}
  S.~P.~Martin,
  Phys.\ Rev.\ D {\bf 70} (2004) 016005
  [hep-ph/0312092].  
  
\bibitem{Martin:2003qz}
  S.~P.~Martin,
  Phys.\ Rev.\ D {\bf 68} (2003) 075002
  [hep-ph/0307101].
  


\end{thebibliography}
\end{document}